\documentclass[12pt]{article}

\usepackage{amscd}
\usepackage{epsfig}
\usepackage{amsmath}
\usepackage{amssymb}
\usepackage{latexsym}                                                       

\newcommand {\newsection}{\setcounter{equation}{0}\section}

\def\graf#1{\begin{center} \epsfig{file=#1, width=10cm}
\end{center}}
\def\grafa#1{\begin{center} \epsfig{file=#1, width=7.7cm}
\quad}
\def\grafb#1{\epsfig{file=#1, width=7.7cm} \end{center}}     

\def\bin#1#2{\pmatrix{ #1 \cr #2 \cr }}
\def\bin#1#2{\left(\begin{array}{c}#1\\#2\end{array}\right)}
\def\f{\frac}
\def\d{\partial}
\def\w{s_{2}}
\def\t{s_{1}}
\def\s{s_{0}}
\def\ux{u_{x}}
\def\uxx{u_{xx}}
\def\uxxx{u_{xxx}}
\def\uxxxx{u_{xxxx}}
\def\uxxxxx{u_{xxxxx}}

\begin{document}


\title{Deformations of bihamiltonian structures of hydrodynamic 
type}
\author{Paolo Lorenzoni}
\maketitle
\begin{center}
\em International School for Advanced Studies (SISSA)
\\via Beirut 2-4, 34014 Trieste, Italy\\{\tt lorenzon@sissa.it}
\end{center}
\vskip 1.5cm
\begin{abstract}

\noindent
In this paper we study the deformations of bihamiltonian PDEs of
hydrodynamic type with one dependent variable.  The reason we study such
deformations is that the deformed systems maintain an infinite number of
commuting integrals of motion up to a certain order in the deformation
parameter.  This fact suggests that these systems could have, at least for
small times, multi-solitons solutions. Our numerical experiments confirm
this hypothesis.

\end{abstract}


\newpage
\newsection{Introduction}

\noindent
The main purpose of this paper is to study the effects of a 
``deformation'' of an infinite dimensional completely integrable 
system.

\noindent
The first point is to define what a deformation is. In this paper we  
deal with 
bihamiltonian systems of hydrodynamic type for which it is 
quite natural to define deformations in terms of the Jacobi identity: the 
deformed bihamiltonian structure satisfies the Jacobi identity only up to 
a certain order in the deformation parameter.

\noindent
The interesting deformations are the deformations that cannot be obtained 
from the original bihamiltonian structure by just a change of coordinates.
Therefore the first problem is to select the non trivial deformations.

\noindent
In order to solve this problem it is convenient formulate it in terms 
of Poisson 
cohomology. 

\noindent
Recently Degiovanni, Magri and Sciacca (see \cite{Magri})
proved that the first two Poisson cohomology groups of a Poisson 
manifold $(M,P)$ are trivial when $M$ is the loop space 
$\{S^{1}\rightarrow {\mathbb R^{n}}\}$ and $P$ is a Poisson bracket of 
hydrodynamic type. Getzler independently (see \cite{Getzler}) proved that 
all groups $H^{i}(P,M)$ for $i$ positive are  trivial for such $(M,P)$.

\noindent
This result, as we will see,  simplifies remarkably our problem and 
allows 
us to solve it (in this paper we classify the deformations up to fourth 
order).

\noindent
For the second order deformations we show explicitly how to obtain an 
infinite ``hierarchy'' of hamiltonian equations . We will see 
that the 
corresponding flows commute up to the order of the deformation.

\noindent
One typical class of solutions of infinite dimensional completely 
integrable systems is the class of multi-soliton solutions. A natural 
question arises: do the equations of the deformed hierarchy have 
multi-solitons solutions (at least for small times)?

\noindent
The numerical experiments we have performed for an equation of the 
deformed hierarchy show the existence of solutions analogous to 
two-solitons solutions.

\noindent
Finally we observe that the deformations of the bihamiltonian structures 
of hydrodynamic type appear in the framework of Frobenius manifolds 
where, with some additional constraints, they play 
a crucial role in the problem of reconstruction of a 2D TFT from a given 
Frobenius manifold studied by Dubrovin and Zhang (see \cite{Zhang}).
One of these constraints, called quasi-triviality, is analysed in the 
last part of this paper.\\  

\noindent
The paper is organized as follows.

\noindent
The first part (section \ref{sec1}) is a brief 
introduction to Poisson cohomology. We focus our attention, in 
particular, on the infinite-dimensional version of Poisson cohomology. 
To do this we use the formalism of formal calculus of variations (see 
for example \cite{Gelfand}). 
The main purpose of section \ref{sec1} is to explain how to get the 
formulae 
for the Schouten brackets that will appear in the calculations.

\noindent
In section \ref{sec2} we  give the 
classification of deformations up to fourth order in the 
deformation parameter. 

\noindent
In section \ref{sec3} we construct the deformed hierarchy and we 
find one 
soliton solutions of one equation of the hierarchy.

\noindent
Section \ref{sec4} gives 
the results of the numerical experiments and section \ref{sec5}  
gives the proof of classification theorem.

\noindent
In the last section (section \ref{sec6}) we introduce the notion of 
quasi-triviality and we prove that all deformations are quasi-trivial.

\newsection{Poisson geometry}
\label{sec1}
\subsection{Poisson bracket}

\newtheorem{de}{Definition}
\begin{de}
Let $M$ be a smooth manifold,  $P^{ij}(x)$ a bivector (i.e. 
a skew-symmetric contravariant 
tensor field of type (2,0)),and $f$ and $g$  two  smooth functions. The 
expression 
\begin{equation}
\{f,g\}:=P^{ij}\frac{\partial f}{\partial x^{i}}\frac{\partial f}{\partial 
x^{j}} 
\end{equation}

\noindent
is a Poisson bracket on $M$ if it defines a structure of Lie algebra
on the ring of smooth functions on $M$.
\end{de}

\noindent
Jacoby identity is the only property of a Lie algebra that is not a 
consequence of the skew-symmetry of $P^{ij}$ and of the definition of 
Poisson 
bracket.  

\noindent
It is well known that the Jacobi identity holds if and only if the tensor
\begin{equation} 
J^{ijk}=\{\{x^{i},x^{j}\},x^{k}\}+cyclic=
\frac{\partial P^{ij}}{\partial x^{s}}P^{sk}+
\frac{\partial P^{jk}}{\partial x^{s}}P^{si}+ 
\frac{\partial P^{ki}}{\partial x^{s}}P^{sj}
\end{equation}

\noindent 
is indentically equal to $0$. In fact 
\begin{equation}
\{\{f,g\},h\}+cyclic=J^{ijk}\frac{\partial f}{\partial x^{i}} 
\frac{\partial g}{\partial x^{j}}\frac{\partial h}{\partial x^{k}}  
\end{equation}

\noindent
for any $f$, $g$ and $h\in C^{\infty}(M)$

\noindent 
\begin{de}
A vector field $\nabla H$ associated to a smooth function $H$ by the 
formula 
\begin{equation}
\left(\nabla H\right)^{i}=P^{ij}\frac{\partial H}{\partial 
x^{j}}=\{x^{i},H\}
\end{equation}

\noindent
is called {\bf Hamiltonian vector field}.
\end{de}

\begin{de}
A smooth function $f$ is called {\bf Casimir} if $\{f,.\}=0$.
\end{de}

\subsection{Schouten bracket and Poisson cohomology}

\noindent
Let $\Gamma^{i}(M)$ be the space of $i$-vectors. It is well-known (see 
\cite{Vaisman}) that there is an unique well defined R-bilinear  
extension of the Lie-derivative to an operator 
\begin{equation*}
\left[.,.\right]:\Gamma^{p}(M)\times\Gamma^{q}(M)\rightarrow\Gamma^{p+q}(M)
\end{equation*}

\noindent
such that
\begin{equation}
\label{wedge}
\left[X_{1}\wedge...\wedge 
X_{p},Q\right]=\sum_{i=1}^{p}(-1)^{i+1}X_{1}\wedge
...\wedge\hat{X}_{i}\wedge...\wedge X_{p}\wedge\left[X_{i},Q\right]
\end{equation}

\noindent
where $X_{k}\in\Gamma^{1}(M)$ for $k=1,..,p$, $Q\in\Gamma^{q}(M)$ and 
$\left[X_{i},Q\right]=L_{X_{i}}Q$.

\noindent
This bilinear map is called {\bf Schouten bracket}.
It has the following properties:
\begin{eqnarray}
\label{first} 
&&\left[P,Q\right]=(-1)^{pq}\left[Q,P\right]\\
&&\left[P,Q\wedge R\right]=\left[P,Q\right]\wedge 
R+(-1)^{pq+q}Q\wedge\left[P,R\right]\\
&&(-1)^{p(r-1)}\left[P,\left[Q,R\right]\right]+
(-1)^{q(p-1)}\left[Q,\left[R,P\right]\right]+
(-1)^{r(q-1)}\left[R,\left[P,Q\right]\right]=0
\end{eqnarray} 
 
\noindent
where $P\in\Gamma^{p}(M)$, $Q\in\Gamma^{q}(M)$ and $R\in\Gamma^{r}(M)$.

\noindent
The last property is called graded Jacobi identity.

\noindent
It can be proved by using (\ref{wedge}) and (\ref{first}). 

\noindent
In coordinates the Schouten bracket of a $p$-vector $P$ and a 
$q$-vector $Q$ is given by the formula (see 
\cite{Vaisman})

\begin{eqnarray*}
&&[P,Q]^{k_{1}...k_{p+q-1}}=\\
&&\f{(-1)^{p}}{p!(q-1)!}\delta^{k_{1}......
k_{p+q-1}}_{i_{1}...i_{p}j_{2}...j_{q}}Q^{uj_{2}...j_{q}}\f{\d
P^{i_{1}...i_{p}}}{\d
x^{u}}+\f{1}{(p-1)!q!}\delta^{k_{1}......k_{p+q-1}}
_{i_{2}...i_{p}j_{1}...j_{q}}P^{ui_{2}...i_{p}}\f{\d Q^{j_{1}...j_{q}}}{\d
x^{u}},
\label{schouten}
\end{eqnarray*}                                                          

\noindent
where $\delta^{...}_{...}$ is the Kronecker multi-index.

\noindent
\newtheorem{ex}{Example}
\begin{ex}
If $P,Q\in\Gamma^{2}(M)$ then
\begin{equation}
\left[P,Q\right]^{ijk}=\frac{\partial Q^{ij}}{\partial 
x^{s}}P^{sk}+\frac{\partial P^{ij}}{\partial x^{s}}Q^{sk}+cyclic
\end{equation}
\end{ex}

\noindent
When $Q=P$ we obtain
\begin{equation}
\frac{1}{2}\left[P,P\right]=J^{ijk}
\end{equation}

\noindent
and then we have the following well-known
\newtheorem{te}{Theorem}
\begin{te}
The Jacobi identity holds if and only if $\left[P,P\right]=0$.

\end{te}

\noindent
The last step before introducing the Poisson cohomolgy is the following 
\begin{te}
Let $P$ be a Poisson bivector on $M$, then the operator 
$d_{P}:\Gamma^{q}(M)\rightarrow\Gamma^{q+1}(M)$ defined by the formula
\begin{equation}
d_{P}Q:=\left[P,Q\right]
\end{equation}

\noindent
is a cohomology operator, i.e. $d_{P}^{2}=0$
\end{te}

\noindent
{\bf Proof}

\noindent
The graded Jacobi identity and the property (\ref{first}) imply
\begin{eqnarray}
0=\left[\left[P,Q\right],P\right]+\left[\left[P,P\right],Q\right]+
\left[\left[Q,P\right],P\right]=2\left[\left[P,Q\right],P\right]=-2d_{P}^{2}Q
\end{eqnarray}

\noindent
The last theorem allows us, following \cite{Lichnerowicz}, to introduce 
the complex
\begin{equation*}
0\rightarrow\Gamma^{0}(M)\rightarrow\Gamma^{1}(M) \rightarrow\Gamma^{2}(M) 
\rightarrow\Gamma^{3}(M) ...
\end{equation*}

\noindent
and to define the Poisson cohomology as 
$HP^{*}(M,P)=ker(d_{P})/Im(d_{P})$.

\noindent
\begin{ex}
$HP^{0}$

\noindent
If $f$ is a smooth function then $d_{P}f=P^{is}\frac{\partial f}{\partial 
x^{s}}$, that is {\bf $HP^{0}$={Casimirs}}.
\end{ex}

\begin{ex}                
$HP^{1}$   

\noindent
The cocycles are the infinitesimal symmetries ($d_{P}X=L_{X}P$), the 
coboundaries are hamiltonian vector fields. Then 
{\bf $HP^{1}$={Infinitesimal 
symmetries}/{Hamiltonian vector fields}}.
\end{ex}

\begin{ex}                
$HP^{2}$
 
\noindent
Let $Q$ be a bivector. $Q$ is a cocycle if and only if 
$\left[P,Q\right]=0$. Therefore $P+\epsilon Q$ satisfies Jacoby identity 
$mod(O(\epsilon^{2}))$. This means that the cocycles are infinitesimal 
deformations of the Poisson bracket. The coboundaries are infinitesimal 
deformations obtained by a change of coordinates. In fact $Q$ is a 
coboundary if and only if $Q=L_{X}P$ where $X$ is a vector field.

\noindent
Summarizing : 

\noindent
{\bf $HP^{2}$={Infinitesimal deformations of 
$P$}/{Deformations obtained by a change of coordinates}}
\end{ex} 

\subsection{Bihamiltonian structure on $M$}

\begin{de}
A bihamiltonian structure on $M$ is a pair $(P_{1},P_{2})$ of Poisson 
bivectors such that for any $\lambda_{1}, \lambda_{2}\in R$, the linear 
combination 
\begin{equation*}
\lambda_{1}P_{1}+\lambda_{2}P_{2}
\end{equation*}

\noindent
is again a Poisson bivector
\end{de}

\noindent
In terms of Schouten bracket the bivectors $(P_{1},P_{2})$ are a 
bihamiltonian structure if and only if
\begin{equation}
[P_{1},P_{1}] = [P_{2},P_{2}] = [P_{1},P_{2}] = 0
\end{equation}

\subsection{Poisson brackets on formal loop space}

\noindent
Now we want to extend the previous definitions to the loop space 
$\mathcal{L}=\{u:S^{1}\rightarrow \mathbb{R}\}$.
Let $\mathcal{A}$ be the space of differential polynomials in $u$, that is
\begin{equation}
f\in \mathcal{A}\Leftrightarrow f=\sum 
f_{s_{1}...s_{m}}(u)u^{(s_{1})}...u^{(s_{m})},
\end{equation}

\noindent
where $u^{(s_{i})}:=\f{d^{s_{i}} u}{d x^{s_{i}}}$.

\noindent
We observe that $f$ is not necessarily a polynomial in $u$ 
\footnote{In general $f$ could depend on $x$. In this paper we
will deal only with differential polynomials where $x$ doesn't appear 
explicitely.}.       

\noindent 
The role of the functions on $\mathcal{L}$ is played by the local 
functionals
\begin{equation*}
I=\int_{s^{1}}f(u(x),u_{x},u_{xx},...)dx
\end{equation*} 

\noindent
where $f\in \mathcal{A}$.
  
\noindent
\begin{de}
A (non local) multivector $X$ is a formal infinite sum of the type
\begin{eqnarray*}
X=X^{s_{1},...,s_{k}}(x_{1},...,x_{k};u(x_{1}),...,u(x_{k}),u_{x}(x_{1}),...)
\frac{\partial}{\partial 
u^{s_{1}}(x_{1})}\wedge...\wedge
\frac{\partial}{\partial u^{s_{k}}(x_{k})}
\end{eqnarray*}

\noindent
where the coefficients satisfy the skew-symmetry condition with respect to 
simultaneous permutations
\begin{eqnarray*}
s_{p},x_{p}\leftrightarrow s_{q},x_{q}
\end{eqnarray*}
\end{de}

\noindent
The wedge product of a $k$-vector $X$ by a $l$-vector $Y$ is defined as 
 
\begin{eqnarray*}
&&(X\wedge 
Y)^{s_{1},...,s_{k+l}}(x_{1},...,x_{k+l};u(x_{1}),...,u(x_{k+l}),...)
=\\
&&\f{1}{k!l!}\sum_{\sigma\in S_{k+l}}(-1)^{sgn \sigma}
X^{s_{\sigma(1)},...,s_{\sigma(k)}}(x_{\sigma(1)},...,x_{\sigma(k)},...)
Y^{s_{\sigma(k+1)},...,s_{\sigma(k+l)}}
(x_{\sigma(k+1)},...,x_{\sigma(k+l)},...).
\end{eqnarray*} 

\begin{de}
A k-vector is called {\bf translation invariant} if
\begin{eqnarray*}
X^{s_{1},...,s_{k}}(x_{1},...,x_{k};u(x_{1}),...,u(x_{k}),...)=
\d_{x_{1}}^{s_{1}}...\d_{x_{k}}^{s_{k}}
X(x_{1},...,x_{k};u(x_{1}),...,u(x_{k}),...) 
\end{eqnarray*}  

\noindent
where $X(...)$ means $X^{0...0}(...)$ and for any $t$

\begin{eqnarray*}
X(x_{1}+t,...,x_{k}+t;u(x_{1}),...,u(x_{k}),...)=
X(x_{1},...,x_{k};u(x_{1}),...,u(x_{k}),...) 
\end{eqnarray*}

\end{de}

\noindent
It follows from this definition that a translation invariant multi-vector 
field is completely characterized by the ``components'' 

\begin{equation*}
X^{x_{1}...x_{k}}:=X(x_{1},...,x_{k};u(x_{1}),...,u(x_{k}),...)
\end{equation*}

\begin{de}
A $k$ form $\omega$ is a finite sum
\begin{equation}
\omega=\f{1}{k!}\omega_{s_{1}...s_{k}}\delta u^{s_{1}}\wedge
...\wedge\delta u^{s_{k}}
\end{equation}

\noindent
where $\omega_{s_{1}...s_{k}}\in \mathcal{A}$
\end{de}

\noindent
In order to define a Poisson structure we need to introduce a criterion 
for the Jacobi identity.

\noindent
We have seen that in the finite-dimensional case the Jacobi identity can 
be written in terms of the Schouten bracket. 

\noindent
Therefore if we will be able to define 
an infinite-dimensional version of the Schouten bracket we will be also 
able to define an infinite-dimensional version of the  Poisson bracket.

\subsubsection{Schouten bracket of translation-invariant multivectors}

\noindent
In the case of translation invariant multivectors one obtains the formula 
for the Schouten bracket simply by translating 
the formula (\ref{schouten}) in the new context.
Heuristically this can be done by substituting sums for integrals, partial 
derivatives for variational derivatives, etc.
The result is:

\begin{de}
Schouten bracket of a translation invariant p-vector 
$P^{x_{1}...x_{p}}$
and a translation invariant q-vector $Q^{x_{1}...x_{q}}$.

\begin{eqnarray}
&&\nonumber [P,Q]^{x_{1}...x_{p+q-1}}=\\
&&\nonumber \sum_{\sigma\in
S_{p+q-1}}(-1)^{sgn(\sigma)}\sum_{s=0}\left(
\sum_{i=1}^{p}\f{(-1)^{p}}{p!(q-1)!}\left(\d^{s}_{x_{\sigma(i)}}
Q^{x_{\sigma(i)}x_{\sigma(p+1)}...x_{\sigma(p+q-1)}}\right)\left(\f{\d
P^{x_{\sigma(1)}...x_{\sigma(p)}}}{\d
u^{(s)}(x_{\sigma(i)})}\right)+\right.\\
&&\left.
\sum_{i=0}^{q-1}\f{1}{(p-1)!q!}\left(\d^{s}_{x_{\sigma(p+i)}}
P^{x_{\sigma(p+i)}x_{\sigma(1)}...x_{\sigma(p-1)}}\right)\left(\f{\d
Q^{x_{\sigma(p)}...x_{\sigma(p+q-1)}}}{\d
u^{(s)}(x_{\sigma(p+i)})}\right)\right)
\label{formula4}
\end{eqnarray}

\noindent
In the case p=2:

\begin{eqnarray}
&&\nonumber
[P,Q]^{x_{1}...x_{q+1}}=\\
&&\nonumber
\sum_{\sigma\in
S_{q+1}}(-1)^{sgn(\sigma)}\sum_{s=0}\left(
\sum_{i=1}^{p}\f{1}{2!(q-1)!}\left(\d^{s}_{x_{\sigma(i)}}
Q^{x_{\sigma(i)}x_{\sigma(3)}...x_{\sigma(q+1)}}\right)\left(\f{\d
P^{x_{\sigma(1)}x_{\sigma(2)}}}{\d
u^{(s)}(x_{\sigma(i)})}\right)+\right.\\
&&\left.
\sum_{i=0}^{q-1}\f{1}{q!}\left(\d^{s}_{x_{\sigma(2+i)}}
P^{x_{\sigma(2+i)}x_{\sigma(1)}}\right)\left(\f{\d
P^{x_{\sigma(2)}...x_{\sigma(q+1)}}}{\d
u^{(s)}(x_{\sigma(2+i)})}\right)\right)
\label{formula5}
\end{eqnarray}
\end{de}                                                 

\noindent

\begin{ex}
Schouten bracket of a bivector $P$ with the components 
$P^{xy}$ and a local functional 
$I=\int_{S^{1}}f(u(x),u_{x},u_{xx},...)dx$

\begin{eqnarray}
[P,I]^{x}=\int_{S^{1}}\f{\delta f}{\delta u(y)}P_{yx}dy
\label{formula1}
\end{eqnarray}

\end{ex}

\begin{ex}
Lie derivative of a translation invariant bivector $P$ with the components
$P^{xy}$
along a translation invariant vector field $Q$ with the components 
$Q^{x}$ 

\begin{eqnarray}
&&\nonumber [P,Q]^{xy}=\\
&&\sum_{s}\left(
\left(\d_{x}^{s}Q^{x}\right)\f{\d P^{xy}}{\d u^{s}(x)}+
\left(\d_{y}^{s}Q^{y}\right)\f{\d P^{xy}}{\d u^{s}(y)}
+\left(\d_{y}^{s}P^{yx}\right)\f{\d Q^{y}}{\d u^{s}(y)}
-\left(\d_{x}^{s}P^{xy}\right)\f{\d Q^{x}}{\d u^{s}(x)}\right)
\label{formula2}
\end{eqnarray}                                                                

\end{ex}
                                                                 
\begin{ex}
Schouten bracket of two translation invariant bivectors $P$ and $Q$

\begin{eqnarray}
&&\nonumber [P,Q]^{xyz}=\f{1}{2}\sum_{s}\left(
\f{\d P^{xy}}{\d u^{s}(x)}\d_{x}^{s}Q^{xz}
+\f{\d Q^{xy}}{\d u^{s}(x)}\d_{x}^{s}P^{xz}
+\f{\d P^{xy}}{\d u^{s}(y)}\d_{y}^{s}Q^{yz}
+\f{\d Q^{xy}}{\d u^{s}(y)}\d_{y}^{s}P^{yz}\right.\\
&&\nonumber \left.+\f{\d P^{zx}}{\d u^{s}(z)}\d_{z}^{s}Q^{zy}
+\f{\d Q^{zx}}{\d u^{s}(z)}\d_{z}^{s}P^{zy}
+\f{\d P^{zx}}{\d u^{s}(x)}\d_{x}^{s}Q^{xy}
+\f{\d Q^{zx}}{\d u^{s}(x)}\d_{x}^{s}P^{xy}
+\f{\d P^{yz}}{\d u^{s}(y)}\d_{y}^{s}Q^{yx}+\right.\\
&&\left.+\f{\d Q^{yz}}{\d u^{s}(y)}\d_{y}^{s}P^{yx}
+\f{\d P^{yz}}{\d u^{s}(z)}\d_{z}^{s}Q^{zx}
+\f{\d Q^{yz}}{\d u^{s}(z)}\d_{z}^{s}P^{zx}\right)
\label{formula3}
\end{eqnarray}                                    

\end{ex}

\noindent
{\bf Remark 1}

\noindent
The operator $\f{\d}{\d u(x)}$ is the usual partial derivative when it 
acts on functions depending on $x$ while it is, by definition, equal to
$0$ when it acts on functions not depending on $x$. In other words 
\begin{eqnarray*}
&&\f{\d u(x)}{\d u(x)}=1\\
&&\f{\d u(y)}{\d u(x)}=0
\end{eqnarray*}

\noindent
{\bf Remark 2}

\noindent
The formula (\ref{formula2}) can be obtained by the formula 
(\ref{formula5}) for $q=1$ by using skew-symmetry of $P$.\\

\begin{de}
A translation invariant Poisson bivector P is a translation invariant
bivector such that
\begin{equation*}
[P,P]_{Schouten}=0
\end{equation*}
\end{de}

\noindent
As in finite dimensional case a translation-invariant Poisson bivector 
$P^{xy}$ defines a Poisson 
structure on the loop space $\mathcal{L}$. The Poisson bracket of two 
local functionals $I_{1}$ ,$I_{2}$ is given by the formula
\begin{equation}
\{I_{1},I_{2}\}:=\int_{S^{1}}\int_{S^{1}}\f{\delta I_{1}}{\delta 
u(x)}P^{xy}\f{\delta I_{2}}{\delta u(y)}dxdy
\end{equation}

\noindent
To define the Poisson cohomology on the space of translation invariant 
multivectors we have to prove that the operator 
$d_{P}:=[P,.]$ associated to a Poisson  bivector $P$
 satisfies the condition $d_{P}^{2}=0$.
We have seen that this condition is satisfied (for a Poisson 
bivector) if 
the graded Jacobi identity holds and moreover that the graded 
Jacobi identity
follows from (\ref{wedge}) and (\ref{first}).

\noindent
The property (\ref{first}) is obvious.
As far as it concerns (\ref{wedge}) it will be sufficient to analyse a 
particular case to understand how the proof works in general.

\noindent
We want to check the ``Leibniz rule'' (\ref{wedge}) when $p=2$ and $Q$ is 
a 
bivector. In this case we have to prove:

\begin{equation}
[X\wedge Y,Q]=Y\wedge[X,Q]-X\wedge[Y,Q]
\end{equation}

\noindent
{\bf Left hand side}

\noindent
By using formula for the Schouten bracket of two bivector, we obtain
\begin{eqnarray*}
&&[X\wedge Y,Q]_{xyz} =\\
&&\left(\f{\d X(x)}{\d u^{s}(x)}Y(y)-X(y)\f{\d Y(x)}{\d u^{s}(x)} 
\right)\left(\d_{x}^{s}P_{xz}\right)+\left(\f{\d P_{xy}}{\d u^{s}(x)}
\right)\left(\left(\d_{x}^{s}X\right)Y(z)-X(z)\left(\d_{x}^{s}X\right)\right)\\
&&\left(\f{\d Y(x)}{\d u^{s}(x)}X(y)-Y(y)
\f{\d X(x)}{\d u^{s}(x)}\right)\left(\d_{y}^{s}P_{yz}\right)
+\left(\f{\d P_{xy}}{\d u^{s}(x)}
\right)\left(\left(\d_{y}^{s}X\right)Y(z)-X(z)\left(\d_{y}^{s}X\right)\right)\\
&&\left(\f{\d X(z)}{\d u^{s}(z)}Y(x)-X(x)
\f{\d Y(z)}{\d u^{s}(z)}\right)\left(\d_{z}^{s}P_{zy}\right)
+\left(\f{\d P_{zx}}{\d u^{s}(z)}
\right)\left(\left(\d_{z}^{s}X\right)Y(y)-X(y)\left(\d_{z}^{s}X\right)\right)\\
&&\left(\f{\d Y(x)}{\d u^{s}(x)}X(z)-Y(z)
\f{\d X(x)}{\d u^{s}(x)}\right)\left(\d_{x}^{s}P_{xy}\right)
+\left(\f{\d P_{zx}}{\d u^{s}(x)}
\right)\left(\left(\d_{x}^{s}X\right)Y(y)-X(y)\left(\d_{x}^{s}X\right)\right)\\
&&\left(\f{\d X(y)}{\d u^{s}(y)}Y(z)-X(z)
\f{\d Y(y)}{\d u^{s}(y)}\right)\left(\d_{y}^{s}P_{yx}\right)
+\left(\f{\d P_{yz}}{\d u^{s}(y)}
\right)\left(\left(\d_{y}^{s}X\right)Y(x)-X(x)\left(\d_{y}^{s}X\right)\right)\\
&&\left(\f{\d Y(z)}{\d u^{s}(z)}X(y)-Y(y)
\f{\d X(z)}{\d u^{s}(z)}\right)\left(\d_{z}^{s}P_{zx}\right)
+\left(\f{\d P_{yz}}{\d u^{s}(z)}
\right)\left(\left(\d_{z}^{s}X\right)Y(x)-X(x)\left(\d_{z}^{s}X\right)\right)\\
\end{eqnarray*}

\noindent
This formula can be written collecting terms in $Y(x)$, $Y(y)$, etc.:

\begin{equation}
=Y(x)\left(-\f{\d X(y)}{\d 
u^{s}(y)}\left(\d_{y}^{s}P_{yz}\right)+\f{\d X(z)}{\d 
u^{s}(z)}\left(\d_{z}^{s}P_{zy}\right)
+\f{\d P_{yz}}{\d u^{s}(y)}\left(\d_{y}^{s}X(y)\right)
+\f{\d P_{yz}}{\d u^{s}(z)}\left(\d_{z}^{s}X(z)\right)\right)
+...
\label{leibniz}
\end{equation}

\noindent
{\bf Right hand side}

\begin{eqnarray*}
&&Y\wedge[X,Q]-X\wedge[Y,Q]=\\
&&=\f{1}{2}\sum_{\sigma\in S_{3}}(-1)^{sgn 
(\sigma)}Y(x_{\sigma(1)})[X,P]_{x_{\sigma(2)}x_{\sigma(3)}} 
-\f{1}{2}\sum_{\sigma\in S_{3}}(-1)^{sgn
(\sigma)}X(x_{\sigma(1)})[Y,P]_{x_{\sigma(2)}x_{\sigma(3)}}=\\
&&=\f{1}{2}\left(Y(x)[X,P]_{yz}-Y(x)[X,P]_{zy}-Y(y)[X,P]_{xz}+Y(y)[X,P]_{zx}
+Y(z)[X,P]_{xy}\right.\\
&&\left.-Y(z)[X,P]_{yx}\right)
-\f{1}{2}\left(Y\leftrightarrow X\right)
\end{eqnarray*}                                                           

\noindent
Let us consider the first two terms:
\begin{eqnarray*}
\f{1}{2}\left(Y(x)[X,P]_{yz}-Y(x)[X,P]_{zy}\right)
\end{eqnarray*}

\noindent
They can be written as
\begin{eqnarray*}
&&Y(x)\left(\f{\d X(z)}{\d
u^{s}(z)}\left(\d_{z}^{s}P_{zy}\right)
-\f{\d X(y)}{\d u^{s}(y)}\left(\d_{y}^{s}P_{yz}\right)
+\f{1}{2}\f{\d P_{yz}}{\d u^{s}(y)}\left(\d_{y}^{s}X(y)\right)
-\f{1}{2}\f{\d P_{zy}}{\d u^{s}(y)}\left(\d_{y}^{s}X(y)\right)+\right.\\  
&&\left.+\f{1}{2}\f{\d P_{yz}}{\d u^{s}(z)}\left(\d_{z}^{s}X(z)\right)
-\f{1}{2}\f{\d P_{zy}}{\d u^{s}(z)}\left(\d_{z}^{s}X(z)\right)\right)
\end{eqnarray*}

\noindent
By comparing this expression with (\ref{leibniz}) we realize that they 
 coincide because of the skew-symmetry:
\begin{eqnarray}
&&\f{\d P_{zy}}{\d u^{s}(y)}=-\f{\d P_{yz}}{\d u^{s}(y)}\\
&&\f{\d P_{zy}}{\d u^{s}(z)}=-\f{\d P_{yz}}{\d u^{s}(z)}
\end{eqnarray}

\noindent
Analogously for the other terms of the left and right hand side.

\subsubsection{Local multivectors and their Poisson cohomology}

\begin{de}
A local k-vector is a traslation invariant k-vector such that its 
dependence on $x_{1}$,...,$x_{k}$ is given by a finite order distribution 
with the support on the diagonal $x_{1}=...=x_{k}$.
\end{de}

\noindent
In coordinates a multivector $X$ has the form
\begin{equation}
X=\sum_{p_{2},p_{3},...,p_{k}\geq 0 
}X(u(x_{1}),u_{x}(x_{1}),...)\delta^{(p_{2})}(x_{1}-x_{2})...
\delta^{(p_{k})}(x_{1}-x_{k})
\end{equation}

\noindent
It is easy to check that $P_{xy}=u\delta'(x-y)+\f{1}{2}u_{x}\delta(x-y)$ 
is 
a local bivector. In fact
\begin{eqnarray*}
P_{yx}=u(y)\delta'(y-x)+\f{1}{2}u_{y}\delta(y-x)=-u_{x}\delta'(x-y)
-u_{x}\delta(x-y)+\f{1}{2}u_{x}\delta(x-y)=-P_{xy}
\end{eqnarray*}

\noindent
But $\f{\d P_{xy}}{\d u(x)}$ is not equal to $-\f{\d P_{yx}}{\d u(x)}=0$!

\noindent
This problem can be solved by writing $P_{xy}$ in a suitable form.
If we write $P_{xy}$ as $P'_{xy}=P_{xy}=\f{1}{2}(P_{xy}-P_{yx})$ 
then
\begin{equation}
\f{\d P'_{xy}}{\d u(x)}=-\f{\d P'_{yx}}{\d u(x)}
\end{equation}

\noindent
This is true in general: if we want to use the same formulae valid for 
non-local multivectors we have to write the local multivector in a form 
``compatible'' with the operators $\f{\d}{\d u^{s}(x)}$. The practical 
rule 
is to write the multivector $T^{x_{1},...,x_{n}}$ in the form 
\begin{equation*}
\f{1}{n!}\sum_{\sigma}(-1)^{sgn(\sigma)}T^{x_{\sigma(1)},...,x_{\sigma(n)}}
\end{equation*}

\noindent
Analogously to the non-local case the formula (\ref{formula5}) allows us  
 to define  
a cohomology operator $d_{P}$ starting from a local Poisson bivector $P$:
\begin{equation}
0\to \Gamma^{0} \buildrel d_{P} \over \to \Gamma^{1} \buildrel d_{P} 
\over \to \Gamma_{local}^{2} \buildrel d_{P} \over \to 
\Gamma_{local}^{3}\to ...\buildrel d_{P} \over \to 
\Gamma_{local}^{k}...  
\end{equation}

\noindent
where $\Gamma^{0}$ is the space of local functionals, 
$\Gamma^{1}$ is the space of non local 
translation-invariant vector fields and $\Gamma_{local}^{k}$ is the 
space of local translation-invariant $k$-vector fields. It is easy to 
check that the cohomology 
groups we have introduced in this way have the same meaning that the 
cohomology 
groups of finite-dimensional Poisson manifolds (see the examples in the 
section devoted to Poisson cohomology of finite dimensional Poisson 
manifolds).  

\newsection{Deformations of bihamiltonian systems of hydrodynamic type}
\label{sec2}

\noindent
In this paper we deal with Poisson bivectors of the form

\begin{equation}
P^{xy}=\phi(u)\delta^{(1)}(x-y)
+\f{1}{2}(\d_{x}\phi)\delta(x-y)
+\sum_{k}\epsilon^{k}P_{xy}^{[k]}
\label{eq1}
\end{equation}

\noindent
where
\begin{equation}
P_{xy}^{[k]}=\sum_{s=0}^{k+1}A_{k,s}(u,u_{x},...)
\delta^{(k+1-s)}(x-y)
\label{deformation}
\end{equation}

\noindent
where $A_{k,s}$ are differential polynomials.
Introducing the following gradation in the space $\mathcal{A}$ of 
differential 
polynomials:
\begin{eqnarray*}
&&deg(f(u))=0\\
&&deg(u^{(k)})=k
\end{eqnarray*} 

\noindent
we require also that 

\begin{equation}
deg(A_{k,s})=s
\label{degree}
\end{equation}

\noindent
The part of order $0$ is called local Poisson bracket of 
hydrodynamic type (see \cite{Novikov}).  

\noindent
Analogously to the finite dimensional case we have the following 

\begin{de}
A bihamiltonian structure on the loop space $L$ is a pair ($P_{1},P_{2}$) 
of local Poisson bivectors such that for any 
$\lambda_{1},\lambda_{2}\in R$, the linear combination
\begin{equation*}
\lambda_{1} P_{1}+\lambda_{2} P_{2}
\end{equation*}

\noindent
is again a Poisson bivector.
\end{de}

\begin{de}
An evolutionary equation
\begin{equation*}
u_{t}=F(u(x),u_{x},u_{xx},....)
\end{equation*}

\noindent
is called {\bf Hamiltonian} if it can be written in the form
\begin{equation*}
u_{t}=\{u,H\}
\end{equation*}

\noindent
where $H$ is some local functional.
\end{de}

\begin{de}
An evolutionary equation is called {\bf bihamiltonian} if and only if 
it is hamiltonian with respect both Poisson bivectors of a bihamiltonian 
structure.
\end{de}

\begin{ex}

\noindent
Rescaling $x\rightarrow \epsilon x$ the KdV equation
$u_{t}=uu_{x}+u_{xxx}$ one obtains
\begin{eqnarray*}
u_{t}=\epsilon(uu_{x}+\epsilon^{2}u_{xxx})
\end{eqnarray*}

\noindent
One usually introduces slow time variable $t\rightarrow\epsilon t$ to
rewrite the last equation into the form

\begin{eqnarray*}
u_{t}=uu_{x}+\epsilon^{2}u_{xxx}
\end{eqnarray*}

\noindent
This is small dispersion expansion of the KdV equation.
It is a bihamiltonian equation.

\noindent
The first Poisson bivector (Gardner-Zakharov-Faddeev bivector)  

\begin{equation}
P^{xy}=\delta'(x-y)
\label{canonical}
\end{equation}                                                               

\noindent
and the second Poisson bivector (Magri bivector) 

\begin{equation}
\{u(x),u(y)\}_{2}=u\delta'(x-y)+\f{1}{2}u_{x}\delta(x-y)-\epsilon^{2}\delta^{3}
(x-y)
\end{equation}

\noindent
belong to the class of bivectors introduced before.
\end{ex}

\noindent
It is easy to check that the pair $(P_{1},P_{2})$ of the 
Gardner-Zakharov-Faddeev bracket and of the Magri bracket is a 
bihamiltonian 
structure. It is called {\bf Magri bihamiltonian structure}.

\begin{de}
The group of transformation 
\begin{equation}
u\rightarrow\bar{u}=\sum_{k}\epsilon^{k}F_{k}(u,u_{x},...)  
\label{miura}
\end{equation}

\noindent
where $F_{k}\in \mathcal{A}$, $deg(F_{k})=k$ and $\f{\d F_{0}}{\d u}\neq 
0$ is called {\bf Miura group}.
\end{de}
 
\noindent
Degiovanni, Magri and Sciacca (see \cite{Magri}) and 
Getzler (see \cite{Getzler}) solved 
independently the problem of studying the action of the Miura group
on the bracket (\ref{deformation}).

\noindent
More precisely they study the following problem: does it exits an 
element of the Miura group that transforms
the bracket (\ref{deformation}) into the bracket (\ref{canonical})?

\noindent
This problem can be reduced to a cohomological problem.
In fact we have seen that it corresponds to the study of the second group 
of Poisson
cohomology  associated to the bracket (\ref{canonical})
\footnote{in our case the role of infinitesimal change of coordinates is
played by the Miura group}.                                                 

\noindent
{\bf $HP^{2}(P,\Pi)$={Infinitesimal deformations of $P$}/{Deformations 
that can 
be obtained by infinitesimal change of coordinates of the form 
(\ref{miura})}}.

\noindent
This cohomology group is trivial (see \cite{Magri} and \cite{Getzler}).

\noindent
In this paper we deal with deformations of bihamiltonian structure of 
hydrodynamic type.

\begin{de}
\label{def}
A deformation of order k of a
bihamiltonian structure of hydrodynamic type is a pair    
$(P_{1},P_{2})$ of the form (\ref{deformation}) such that
$P_{1}-\lambda P_{2}$ satisfies Jacoby identity for every
$\lambda$ up to the order $k$:
\begin{equation}
[P_{1},P_{1}]=[P_{2},P_{2}]=[P_{1},P_{2}]=o(\epsilon^{k})
\end{equation}
\end{de}
                                                                          
\noindent
From the triviality of second Poisson cohomology group it follows that 
we can always assume one of the  brackets of the form (\ref{canonical}).  

\noindent
In the next section we explain how to classify these deformations.                                            
                                                                                 
\subsection{Classification of deformations of bihamiltonian structure 
of 
hydrodynamic type}

\noindent 
We are interested in the following question: which deformations of a 
bihamiltonian structure of hydrodynamic type are 
trivial? In other words which deformations  can be obtained from a 
bihamiltonian structure of hydrodynamic type by the action of Miura group?

\noindent
We start considering first order 
defomations, that is $P_{1}=P^{(0)}_{1}=\delta^{1}(x-y)$ and $P_{2}
=P^{(0)}_{2}+\epsilon P^{(1)}_{2}$

\noindent
By definition we have\footnote{It is well-known that $[P_{1},P_{1}]=0$}
\begin{eqnarray}
&&[P_{1},P_{2}]=o(\epsilon)\\
&&[P_{2},P_{2}]=o(\epsilon)
\end{eqnarray}

\noindent
The equation $[P_{1},P_{2}]=o(\epsilon)$ implies 
$[P_{1}^{(0)},P_{2}^{(1)}]=d_{1}P^{(1)}_{2}=0$ and from the triviality of 
the second Poisson cohomology group it follows that there exists a 
vector  
$X^{(1)}_{2}$ such that 

\begin{equation}
P^{(1)}_{2}=d_{1}X^{(1)}_{2}
\end{equation}

\noindent
The equation $[P_{2},P_{2}]=o(\epsilon)$ implies that

\begin{equation} 
[P^{(0)}_{2},P^{(1)}_{2}]=d_{2}P_{2}^{(1)}=d_{2}d_{1}X_{2}^{(1)}
=-d_{1}d_{2}X_{2}^{(1)}=0 
\end{equation}

\noindent
where we have used the fact that $(d_{1}+d_{2})^{2}=0$. 

\noindent
Among all deformations that satisfy the equation 
$d_{1}d_{2}X_{2}^{(1)}=0$ we have to select trivial deformations, that is 
the deformations that can be obtained by infinitesimal change of 
coordinates. In our case this means that there exists a vector field 
$\tilde{X}$ such that 

\begin{eqnarray}
&&Lie_{\tilde{X}}P^{(0)}_{1}=0\\
&&Lie_{\tilde{X}}P^{(0)}_{2}=P^{(1)}_{2}
\end{eqnarray}      

\begin{te} 
\label{trivial}
A deformation 
$P_{\lambda}=P^{(0)}_{1}-\lambda(P^{(0)}_{2}+\epsilon d_{1}X^{(1)}_{2})$ 
is trivial $\Leftrightarrow$
$X^{(1)}_{2}=d_{1}a+d_{2}b$ where a, b are 
local functionals.
\end{te} 

\noindent
{\bf Proof}

\noindent
"$\Leftarrow$"

\noindent
$X^{(1)}_{2}=d_{1}a+d_{2}b$ $\Rightarrow$
$P^{(1)}_{2}=d_{1}d_{2}b=
-d_{2}d_{1}b$

\noindent
This means $P^{(1)}_{2}=Lie_{\tilde{X}}P^{(0)}_{2}$ with
$\tilde{X}=-d_{1}b$.

\noindent
Moreover $Lie_{\tilde{X}}P^{(0)}_{1}=0$.

\noindent
"$\Rightarrow$"

\noindent
$Lie_{\tilde{X}}P^{(0)}_{1}=0\Rightarrow\tilde{X}=d_{1}b$ (the
first Poisson cohomology group is trivial).

\noindent
$-d_{1}d_{2}b=d_{2}d_{1}b=Lie_{\tilde{X}}P^{(0)}_{2}=
P^{(2)}_{2}=d_{1}X^{(2)}_{2}$ 
$\Rightarrow$ $X^{(2)}_{2}=-d_{2}b+d_{1}a$
\\

\noindent
From the last theorem it follows that the elements of the 
group  $Ker(d_{1}d_{2})/\left(Im(d_{1})+Im(d_{2})\right)$ are the non 
trivial first order deformations.

\noindent
{\bf Remark}

\noindent
For higher order 
deformations we can repeat the same arguments and obtain 
the same result. Consequently, in general:

\noindent
1) A k order deformation $P^{(k)}_{2}$ can be represented in the form
\begin{equation}
P^{(0)}_{2}-\lambda P^{(0)}_{1}+\sum_{i=1}^{k}\epsilon^{i}P^{(i)}_{2}=
P^{(0)}_{2}-\lambda P^{(0)}_{1}+\sum_{i=1}^{k}\epsilon^{i}d_{1}X^{(i)}_{2}
\end{equation}

\noindent
where due to our definition of a gradation in the 
space $\mathcal{A}$ of differential polynomials (see 
(\ref{degree})) we have necessarily $deg(X^{(i)}_{2})=i$ (see 
 the form of the formula (\ref{formula2}) when 
$P_{1}=\delta'(x-y)$). 

\noindent
2) $P^{(k)}_{2}$ is trivial if and only if 
$X^{(k)}_{2}=d_{1}A+d_{2}B$. 

\noindent
{\bf Remark 1}
Also in this case ``trivial'' means that $P^{(k)}_{2}$ can be eliminated 
by the 
action of Miura group. Obviously the triviality of $P^{(k)}_{2}$ doesn't 
imply that the $k$-order deformation is trivial.\\ 

\noindent
{\bf Remark 2}

\noindent
From the (\ref{formula1}) it follows immediately that $d_{1}$ and $d_{2}$ 
increases the degree of a local functional by one\footnote{we will 
identify the degree of a local functional $\int fdx$ with the degree of 
its density $f$}. 
Then the degree of $A$ and $B$ must be equal to $k-1$.\\ 

\noindent
In the next section we summarize the results of the classification of
deformations up to fourth order.   
                                    
\subsection{Classification of deformations: results} 

\begin{te}
\label{classification}
Up to the fourth order all deformations of a bihamiltonian structure of 
hydrodynamic type (see definition \ref{def}) can be reduced, by the action 
of 
Miura group to the following form:

\begin{eqnarray}
&&\nonumber
 u\delta^{(1)}(x-y)+\f{1}{2}u_{x}\delta(x-y)-\lambda 
\delta^{(1)}(x-y)+\\
&&\nonumber +\epsilon^{2}\left(
-2s\delta^{3}(x-y)-3\d_{x}s\delta^{2}(x-y)
-\d_{x}^{2}s\delta^{1}(x-y)\right)+\\                
&&\nonumber +\epsilon^{4}\left(
-2\tilde{s}\delta^{5}(x-y)
-5\left(\d_{x}\tilde{s}\right)\delta^{4}(x-y)
-10\left(\d^{2}_{x}\tilde{s}\right)\delta^{3}(x-y)
-10\left(\d_{x}^{3}\tilde{s}\right)\delta^{2}(x-y)+\right.\\
&&\left.-3\left(\d_{x}^{4}\tilde{s}\right)\delta^{1}(x-y)
+2w\delta^{3}(x-y)+3\left(\d_{x}w\right)\delta^{2}(x-y)
+\left(\d_{x}^{2}w\right)\delta^{1}(x-y)\right)+O(\epsilon^{5})               
\label{class}
\end{eqnarray}

\noindent
whith 

\begin{equation}
w=2\f{\d\tilde{s}}{\d u}u_{xx}
\end{equation}

\noindent
where $s$ is an arbitrary function of $u$ and 
$\tilde{s}=-2s\f{\d s}{\d u}$.
\end{te}

\noindent
From this theorem it follows immediately:
\newtheorem{co}{Corollary}
\begin{co}
Up to the fourth order every deformation of Magri bihamiltonian structure 
is trivial
\end{co}

\noindent
{\bf Proof}

\noindent
Indeed, in this case $s=constant\neq 0$ and then $\tilde{s}=0$, $w=0$.

\newsection{The deformed hierarchy}
\label{sec3}
\subsection{Integrals of motions}

\noindent
As we have already been said bihamiltonian structures 
give rise to
an infinite number of "almost-constants" of motions. From these constants
one can construct an hierarchy of hamiltonian equations.

\noindent
In this section we
show explicitely how to produce this hierarchy and we check that
the corresponding flows commute up to the order of the deformation.

\noindent
The technique is well known for the KdV equation (see \cite{Falqui}): one 
looks for
a solution of the equation
\begin{equation}
P_{\lambda}v=0
\end{equation}

\noindent
in terms of a formal series:
the coefficients of 1-form $v$ are
variational derivatives of the integrals of motions. 
\footnote{this is equivalent to the exactness of the 1-form $v$ with 
respect 
to the vertical differential of the variational bicomplex (see 
\cite{Dedecker}).}

\noindent
We apply the same technique to the second order deformations (see 
classification theorem). 

\noindent
We start with the equation $P_{\lambda}v=0$. This equation implies
$v P_{\lambda}v=0$. From skew-symmetry of $P_{\lambda}$ it follows that
$0=v P_{\lambda}v=\partial_{x}(...)$. That is $(...)=f(\lambda)$. More
precisely 
\begin{equation}
\label{rel1}
v(-\lambda\partial_{x}+u\partial_{x}+\frac{1}{2}u_{x}+\epsilon^{2}(2s(u)
\partial_{x}^{3}+3(\partial_{x}s)\partial_{x}^{2}+(\partial_{x}^{2}s)
\partial_{x}))v=0
\end{equation}

\noindent
From (\ref{rel1}) it follows:
\begin{eqnarray}
\partial_{x}(\f{1}{2}v^{2}(u-\lambda)+\epsilon^{2}(2svv_{xx}+s_{x}vv_{x}
-sv_{x}^{2}))=0
\end{eqnarray}

\noindent
that is
\begin{eqnarray}
\f{1}{2}v^{2}(u-\lambda)+\epsilon^{2}(2svv_{xx}+s_{x}vv_{x}-sv_{x}^{2})
=f(\lambda)
\end{eqnarray}

\noindent
By choosing $f(\lambda)=-\f{\lambda}{2}$ we look for a solution of the
form 
\begin{equation}
v=\sum_{i=0}\f{p_{i}}{\lambda^{i}}
\end{equation}

\noindent
where obviously  $p_{0}=1$.

\noindent
By straightforward calculation we get
\begin{equation}
\f{1}{2}\sum_{i+j=k}\f{p_{i}p_{j}}{\lambda^{k-1}}=\sum_{i+j=k-1}(\f{u}{2}
p_{i}p_{j}+\epsilon^{2}(2sp_{i}p_{jxx}+s_{x}p_{i}p_{jx}
-sp_{ix}p_{jx})\f{1}{\lambda^{k-1}}
\label{rel2}
\end{equation}

\noindent
for $k=1$ (\ref{rel2}) implies
\begin{equation}
\f{1}{2}(p_{0}p_{1}+p_{1}p_{0})=\f{1}{2}up_{0}^{2}
\end{equation}

\noindent
that is  
\begin{equation}
p_{1}=\f{u}{2}
\end{equation}

\noindent
for k=2:                                                       
\begin{equation}
p_{2}=\f{3}{8}u^{2}+\f{\epsilon^{2}}{2}(2su_{xx}+s_{x}u_{x})
\end{equation}

\noindent
and so on. We observe that the coefficients will have always the form 
\begin{equation}
p_{i}=f(u)+O(\epsilon^{2})+O(\epsilon^{4})+...
\label{formula6}
\end{equation} 

\noindent
where $f(u)$ is a
polynomial in u.

\noindent
Now we want to prove that the coefficients $p_{i}$ are, up to the 
second order variational derivatives. We consider a curve $u(t)$ on the 
loop space $L$. By differentiating the equation

\begin{equation}
\f{1}{2}v^{2}(u-\lambda)+\epsilon^{2}(2svv_{xx}+s_{x}vv_{x}-sv_{x}^{2})
+\f{\lambda}{2}=0
\label{rel4}
\end{equation}

\noindent
along the vector field $\dot{u}$, tangent to this curve, we get 
\begin{eqnarray*}
&&\dot{u}v=-2\dot{v}(u-\lambda)-\epsilon^{2}\left(4\dot{s}v_{xx}
+4s\f{\dot{v}}{v}v_{xx}+4s\dot{v_{xx}}+2\dot{s_{x}}v_{x}
+2s_{x}\f{\dot{v}}{v}v_{x}+2s_{x}\dot{v_{x}}+\right.\\
&&\left.-2\dot{s}\f{v_{x}^{2}}{v}
-4s\f{v_{x}}{v}\dot{v_{x}}\right)
\end{eqnarray*}

\noindent
By straightforward calculation one can write the last equation in the
form                                                                             

\begin{eqnarray}
\dot{u}v=\d_{t}\left(-2\f{\lambda}{v}\right)+\epsilon^{2}\left(\d_{t}
\left(-4sv_{xx}-2s_{x}v_{x}\right)+\d_{x}\left(4s\f{\dot{v}}{v}v_{x}\right)
+2\f{v_{x}}{v}\left(\dot{s}v_{x}-s_{x}\dot{v}\right)\right)
\end{eqnarray}

\noindent
This equation implies
\begin{equation}
\int_{S^{1}}\dot{u} v dx=\f{d}{dt}
\left(\int_{S^{1}}fdx\right)+O(\epsilon^{4})
\end{equation}

\noindent
for some $f=f(u,u_{x},u_{xx},...)$.

\noindent
In fact from (\ref{formula6}) it follows that $v$ has the form
\begin{equation} 
v=v^{0}+O(\epsilon^{2})
\end{equation}

\noindent
where $v^{0}$ is a function of $u$
and 
\begin{eqnarray*}
\dot{s}v_{x}^{0}-s_{x}\dot{v^{0}}=0
\end{eqnarray*}

\noindent
Now we can formulate the

\begin{te}
$p_{i}=\f{\delta I_{i}}{\delta u} + O(\epsilon^{4})$ for some functionals 
$I_{i}$. Moreover
\begin{equation}
\{I_{i},I_{j}\}_{1}=O(\epsilon^{4})
\end{equation}
\end{te}

\noindent
{\bf Proof}
\begin{eqnarray*}
&&\int_{S^{1}}\dot{u} v dx=
\f{d}{dt}\left(\int_{S^{1}}fdx\right)+O(\epsilon^{4})=
\int_{S^{1}}\left(\f{\d f}{\d u}\dot{u}+\f{\d f}{\d u_{x}}\dot{u_{x}}+
\f{\d f}{\d u_{xx}}\dot{u_{xx}}+...\right)dx+O(\epsilon^{4})=...
\end{eqnarray*}

\noindent
By integrating by parts, we get
\begin{eqnarray*}
...=\int_{s^{1}}\dot{u}\left(\f{\d f}{\d u}
-\d_{x}\left(\f{\d f}{\d u_{x}}\right)+
\d_{x}^{2}\left(\f{\d f}{\d u_{xx}}\right)+...\right)dx+O(\epsilon^{4})=
\int_{S^{1}}\dot{u}\f{\delta f}{\delta u}dx+O(\epsilon^{4})
\end{eqnarray*} 

\noindent
Then
\begin{equation}
v=\f{\delta f}{\delta u}+O(\epsilon^{4})
\end{equation}

\noindent 
that is
\begin{equation}
p_{i}=\f{\delta I_{i}}{\delta u}+O(\epsilon^{4}) 
\end{equation}

\noindent
for some functionals $I_{i}$.

\noindent                                
Moreover, by definition, the coefficients $p_{i}$ of 1-form $v$
satisfy the equation 
\begin{equation}
(P_{2}-\lambda P_{1})\left(\sum_{i=0}\f{p_{i}}{\lambda^{i}}\right)=0
\end{equation}

\noindent
In terms of $I_{i}$ this
condition can be written as
\begin{equation}
\sum_{i=0}\left(P_{2}\f{\delta I_{i}}{\delta u}\right)\f{1}{\lambda^{i}}=
\sum_{i=-1}\left(P_{1}\f{\delta I_{i+1}}{\delta 
u}\right)\f{1}{\lambda^{i}}+O(\epsilon^{4})  
\end{equation}

\noindent
From this equation it follow immediately:
\begin{eqnarray*}
&&P_{1}\f{\delta I_{0}}{\delta u}=O(\epsilon^{4})\\
&&...\\
&&P_{2}\f{\delta I_{i}}{\delta u}=P_{1}\f{\delta I_{i+1}}{\delta 
u}+O(\epsilon^{4})\\
&&...
\end{eqnarray*}

\noindent
By using these Lenard recursion relations it is easy to prove the therem 
(see for details \cite{Falqui})
 
\subsection{Soliton solutions}
                                                                            
\noindent
Now we study one of the equation of the hierarchy. More precisely
we concentrate on the equation
\begin{equation}
u_{t}=\d_{x}\f{\delta I_{2}}{\delta
u}=\f{3}{4}uu_{x}+\epsilon^{2}\left(s(u)u_{xxx}+2\f{\d
s}{\d u}u_{x}u_{xx}+\f{1}{2}\f{\d^{2}s}{\d u^{2}}u_{x}^{3}\right)
\label{main}
\end{equation}

\noindent
When $s=constant$ this is KdV equation.

\subsubsection{One soliton solutions}

\noindent
We look for solutions of (\ref{main}) of the form
$u(x,t)=u(x+ct)=u(z)$.

\noindent
By substituting we get
\begin{equation}
cu_{z}=\f{3}{4}uu_{z}+\epsilon^{2}\left(s(u)u_{zzz}+2\f{\d s}{\d
u}u_{z}u_{zz}+\f{1}{2}\f{\d^{2}s}{\d u^{2}}u_{z}^{3}\right)
\end{equation}

\noindent
This equation can be written as
\begin{equation}
\d_{z}\left(cu-\f{3}{8}u^{2}-\epsilon^{2}\left(\d_{z}(su_{z})
-\f{1}{2}(\d_{z}s)u_{z}\right)\right)=0
\end{equation}

\noindent
that is                                                          

\begin{equation}
cu-\f{3}{8}u^{2}-\epsilon^{2}\left(\d_{z}(su_{z})
-\f{1}{2}(\d_{z}s)u_{z}\right)=c_{1}
\end{equation}

\noindent
By multiplying for $u_{z}$ we get again a total derivative 
\begin{equation}
\d_{z}\left(\f{1}{2}cu^{2}-\f{1}{8}u^{3}-\epsilon^{2}\left(\f{1}{2}
su_{z}^{2}\right)-c_{1}u\right)=0
\end{equation}

\noindent
that is
\begin{equation}
\f{1}{2}cu^{2}-\f{1}{8}u^{3}-\epsilon^{2}\left(\f{1}{2}
su_{z}^{2}\right)-c_{1}u=c_{2}
\end{equation}

\noindent
This equation can be written as
\begin{equation}
(\f{du}{dz})^{2}=F(u)
\end{equation}                                                              

\noindent
where
\begin{equation}
F(u)=\f{2}{\epsilon^{2}s(u)}\left(-\f{1}{8}u^{3}+\f{1}{2}cu^{2}
-c_{1}u-c_{2}\right)
\end{equation}

\noindent
To obtain the solution one has to invert the following integral
\begin{equation}
z-z_{0}=\pm\int_{u}^{u_{0}}\f{1}{F(u)^{1/2}}du
\end{equation}

\subsubsection{Case s(u)=u}

\noindent
In this case $F(u)=\f{P(u)}{u}$ where $P(u)$ is a polynomial of degree
3. It is well known (see \cite{Drazin}) that one soliton solutions occur 
when $F(u)$ has
one simple zero and one double zero.

\noindent
By using the formula (see \cite{Prudnikov})
\begin{equation}
\int_{x}^{a_{0}}\left(\f{x-a_{3}}{(a_{0}-x)(x-a_{1})(x-a_{2})}\right)
^{\f{1}{2}}dx=\f{2(a_{0}-a_{3})}
{\left((a_{0}-a_{2})(a_{1}-a_{3})\right)^{\f{1}{2}}}\Pi\left(\phi,
\f{a_{1}-a_{0}}{a_{1}-a_{3}},k\right)
\end{equation}

\noindent
where $a_{0}>u> a_{1}> a_{2} > a_{3}$ and
\begin{eqnarray}
&&\phi=arcsin\left(\f{(a_{1}-a_{3})(a_{0}-x)}{(a_{0}-a_{1})(x-a_{3})}\right)
^{\f{1}{2}}\\
&&k=\left(\f{(a_{0}-a_{1})(a_{2}-a_{3})}{(a_{0}-a_{2})(a_{1}-a_{3})}\right)
^{\f{1}{2}}\\
&&\Pi(\phi,\nu,k)=\int_{0}^{\phi}\f{d\phi}{\left(1-\nu 
sin^{2}\phi\right)\left(1-k^{2}sin^{2}\phi\right)^{\f{1}{2}}}
\end{eqnarray}

\noindent
we obtain, when $a_{1}=a_{2}$ and $a_{3}=0$ (see \cite{Abramowitz})

\begin{eqnarray}
&&z-z_{0}=\\
&&\pm\f{2a_{1}}
{\left(a_{1}\left(a_{0}-a_{1}\right)\right)^{\f{1}{2}}}
\left(
-2ln\left(\f{\left(\f{a_{1}\left(a_{0}-u\right)}{u\left(a_{0}-a_{1}\right)}
\right)^{\f{1}{2}}+1}{\left(\f{a_{0}\left(u-a_{1}\right)}{u\left(a_{0}
-a_{1}\right)}\right)^{\f{1}{2}}}\right)
-\left(\f{a_{0}}{a_{1}}-1\right)^{\f{1}{2}}arctg\left(\f{2u\left(\f{a_{0}}{u}
-1\right)
^{\f{1}{2}}}{2u-a_{0}}\right)\right)
\label{elliptic}
\end{eqnarray}

\noindent
The  speed $c$ of the wave and  the constants of integration $c_{1}$ and 
$c_{2}$ can be easily written in terms of the coefficients $a_{0}$, 
$a_{1}$, $a_{2}$, $a_{3}$:

\begin{eqnarray}
&&c=\f{1}{4}(a_{0}+2a_{1})\\
&&c_{1}=\f{1}{8}(2a_{0}a_{1}+a_{1}^{2})\\
&&c_{2}=-\f{1}{8}a_{0}a_{1}^{2}
\end{eqnarray}

\noindent
The expression (\ref{elliptic}) can be inverted numerically. 
To test the existence of two solitons solutions we have used two such 
solutions with different speed as initial condition. The results of 
numerical experiments are described in the next section.

\newsection{Numerical experiments}
\label{sec4}
\graf{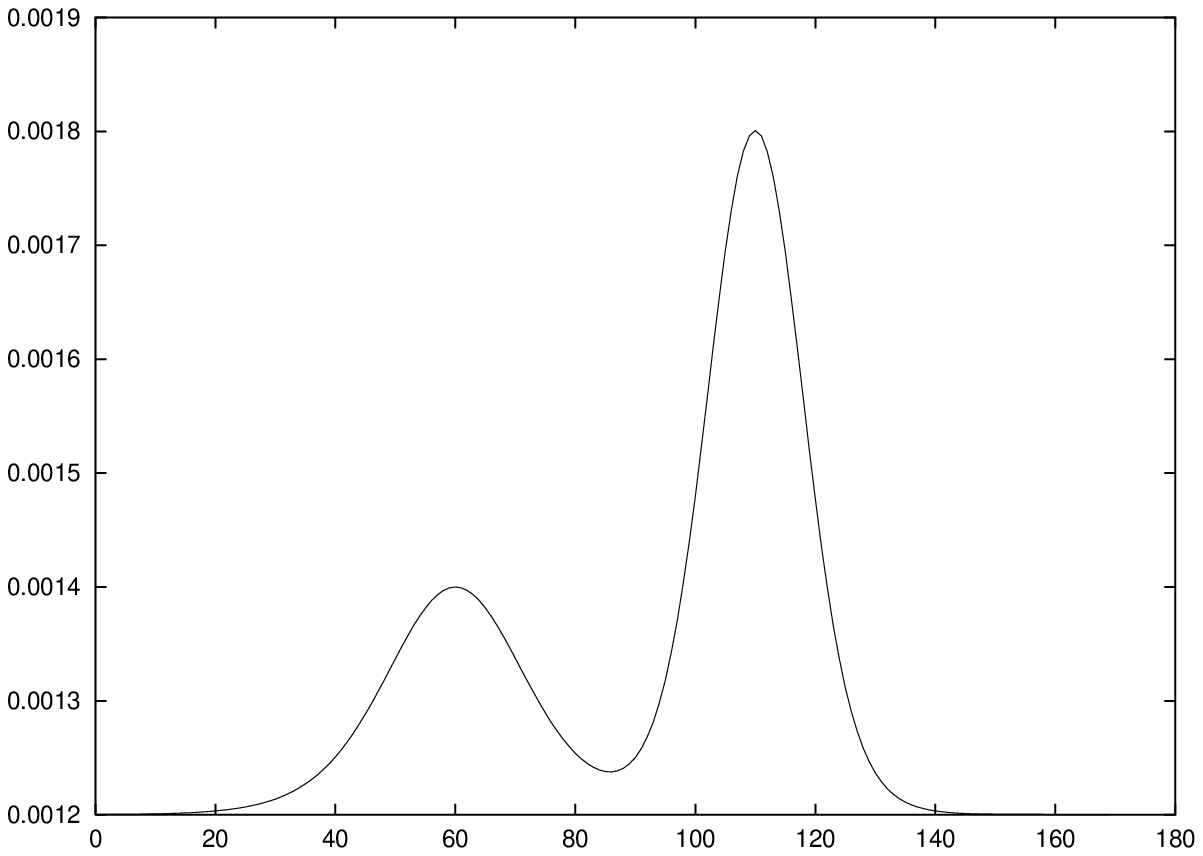} 

\noindent
In this section we analyse the equation (\ref{main}) for $s(u)=u$, i.e.
\begin{equation*}
u_{t}=\d_{x}\left(\f{\delta I_{2}}{\delta 
u}\right)=\d_{x}\left(\f{3}{8}u^{2}+\f{\epsilon^{2}}{2}\left(2uu_{xx}+u_{x}^{2}\right)\right)
\end{equation*}

\noindent
We write this equation in the form
\begin{equation*}
u_{t}+F_{x}=0
\end{equation*}

\noindent
where $F=-\f{\delta I_{2}}{\delta u}$. To make numerical 
experiments we have used a two-steps Lax-Wendroff scheme (see 
\cite{Potter}). This 
scheme is characterized by an auxiliary step of calculation
\begin{equation*}
u^{n+\f{1}{2}}_{j+\f{1}{2}}=\f{1}{2}(u^{n}_{j}+u^{n}_{j+1})-\f{\Delta 
t}{2\Delta x}(F^{n}_{j+1}-F^{n}_{j}),
\end{equation*}

\noindent
where $F^{n}_{j+1}=F(u^{n}_{j+1})$.

\noindent
The main step is
\begin{equation*}
u^{n+1}_{j}=u^{n}_{j}-\f{\Delta
t}{\Delta x}(F^{n+\f{1}{2}}_{j+1+\f{1}{2}}
-F^{n+\f{1}{2}}_{j-\f{1}{2}})                                  
\end{equation*}

\noindent
The first figure illustrates two solitons of different height and 
speed 
we have used in numerical experiments.

\noindent
We have choosen periodic boundary conditions.

\noindent
The results of our experiments is showed in figure 2. 
The waves move from the right to the left.
The three pictures of the second figure show the solitons before, 
during and 
after the collision. Like the soliton collision in integrable systems they 
reemerge with the same shape.     

\graf{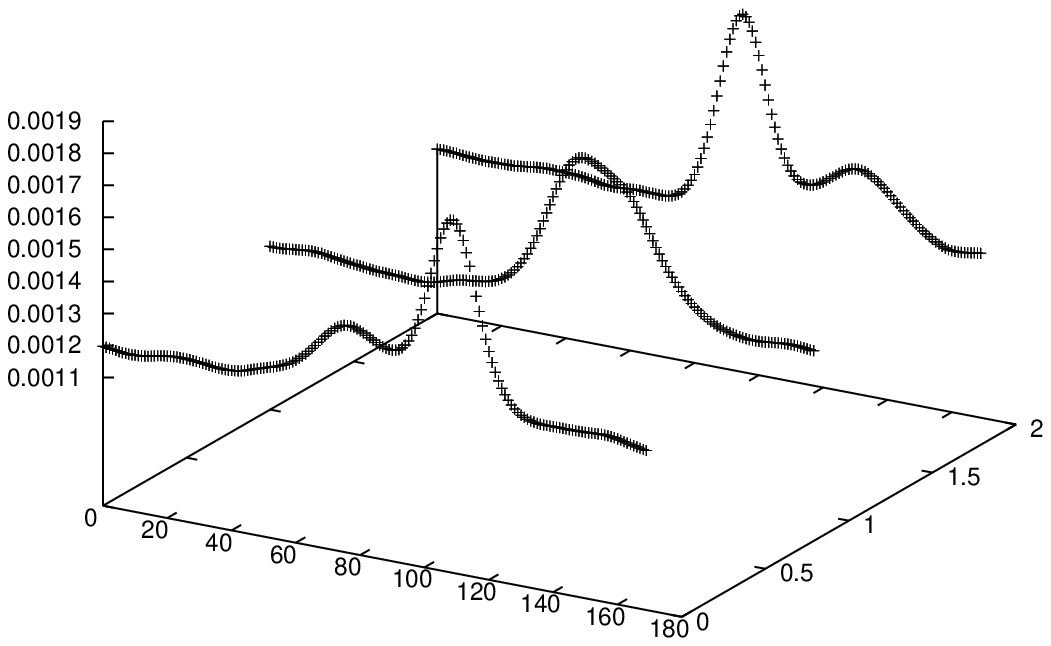}

\noindent
{\bf Remark}

\noindent
Due to numerical instability for big 
amplitude we performed numerical experiments with small and slow 
waves.

\noindent
We don't know if the origin of this numerical instability is the alghoritm 
we have used. 
 
\newsection{Classification of deformations: proof}
\label{sec5}
\subsection{First order deformations}

\noindent
We have seen that in this case the non trivial deformations are the 
elements of the group $Ker(d_{1}d_{2})/\left(Im(d_{1})+Im(d_{2})\right)$.

\noindent
We start looking for the solutions of the equation
$d_{1}d_{2}X=0$.

\noindent
First of all we calculate $d_{2}X$ 

\begin{eqnarray*}
&&d_{2}X=[P_{2}^{(0)},X]_{xy}=\frac{1}{2}
(\partial_{x}^{s}X)\frac{\partial}{\partial
u^{s}(x)}(u\delta'(x-y)+\frac{1}{2}u_{x}\delta(x-y))\\
&&-\frac{1}{2}
(\partial_{y}^{s}X)\frac{\partial}{\partial u^{s}(y)}
(u\partial_{y}\delta(y-x)+\frac{1}{2}u_{y}\delta(y-x))+
\partial^{s}_{y}(u\partial_{y}\delta(y-x)+\frac{1}{2}u_{y}
\delta(y-x))\frac{\partial X}{\partial
u^{s}(y)}\\
&&-\partial_{x}^{s}(u\delta'(x-y)+\frac{1}{2}u_{x}\delta(x-y))
\frac{\partial X}{\partial u^{s}(x)}
\end{eqnarray*}

\noindent
where, by definition,  $\delta'(x-y)=\delta^{(1)}(x-y)=\d_{x}\delta(x-y)$.

\noindent
Moreover by using the formula
\begin{eqnarray}
f(y)\delta^{(s)}(x-y)=\sum_{q=0}^{s}\bin{s}{q}
f^{(q)}(x)\delta^{(s-q)}(x-y)
\label{formula}
\end{eqnarray}                                 

\noindent
we get 
\begin{eqnarray*}
&&\f{1}{2}\left(
\frac{\partial}{\partial u^{s}(x)}
(u\delta'(x-y)+\frac{1}{2}u_{x}\delta(x-y))(\partial^{s}_{x}X)-  
\frac{\partial}{\partial u^{s}(y)}
(u\partial_{y}\delta(y-x)+\frac{1}{2}u_{y}\delta(y-x))(\partial^{s}_{y}X)
\right)\\
&&=\f{1}{2}\left(X\delta'(x-y)+\frac{1}{2}\partial_{x}X\delta(x-y)-
X\partial_{y}\delta(x-y)-\frac{1}{2}\partial_{y}X\delta(y-x)\right)=\\ 
&&=X\delta'(x-y)+\f{1}{2}\partial_{x}X\delta(x-y)
\end{eqnarray*}

\noindent
and
\begin{eqnarray*}
&&\partial_{y}^{s}(u\partial_{y}\delta(y-x)+\frac{1}{2}u_{y}\delta(y-x))\frac{\partial
X}{\partial
u^{s}(y)}-\partial_{x}^{s}(u\delta^{(1)}(x-y)+\frac{1}{2}u_{x}\delta(x-y))
\frac{\partial X}{\partial u^{s}(x)}=\\
&&=\partial_{y}^{s}(-u(x)\delta^{(1)}(x-y)-\frac{1}{2}u_{x}\delta(x-y))
\frac{\partial X}{\partial
u^{s}(y)}+\\
&&-\partial_{x}^{s}(-u(y)\partial_{y}\delta(y-x)-\frac{1}{2}u_{y}
\delta(y-x))\frac{\partial X}{\partial u^{s}(x)}=\\
&&(-1)^{s+1}(u(x)\delta^{(s+1)}(x-y)+\frac{1}{2}u_{x}\delta^{(s)}(x-y))
\frac{\partial X}{\partial
u^{s}(y)}+\\
&&-u(y)\delta^{(s+1)}(x-y)+\frac{1}{2}u_{y}\delta^{(s)}(x-y))
\frac{\partial X}{\partial u^{s}(x)}
\label{abel}
\end{eqnarray*}

\noindent
Summarizing we obtain
\begin{eqnarray}
(d_{2}X)_{xy}=X\delta^{(1)}(x-y)+\frac{1}{2}\partial_{x}X\delta(x-y)+
\sum_{s}\sum_{q=0}^{s+1}c_{qs}\delta^{(s+1-q)}(x-y)
\label{arnold}
\end{eqnarray}

\noindent
with
\begin{eqnarray*}
c_{qs}=\bin{s+1}{q}\left(\left(-1\right)^{s+1}u\left(\partial^{q}_{x}
\left(\frac{\partial X}{\partial u^{s}(x)}\right)\right)-\frac{\partial
X}{\partial u^{s}(x)}u^{(q)}\right)
\end{eqnarray*}
\begin{eqnarray}
+\frac{1}{2}\bin{s}{q-1}\left(\left(-1\right)^{s+1}u_{x}
\left(\partial_{x}^{q-1}\left(\frac{\partial X}{\partial
u^{s}(x)}\right)\right)+\frac{\partial X}{\partial u^{s}(x)}u^{(q
)}\right)
\end{eqnarray}

\noindent
In this case $deg(X)=1$ i.e. $X=s(u)u_{x}$.
This implies that we can write the sum (\ref{arnold}) as 
\begin{eqnarray}
&&\nonumber  (d_{2}X)_{xy}=\\
&&\left(\frac{1}{2}\partial_{x}X+c_{21}+c_{10}\right)\delta(x-y)+
\left(X+c_{11}+c_{00}\right)\delta^{(1)}(x-y)+c_{01}\delta^{(2)}(x-y)
\label{silov}
\end{eqnarray}

\noindent
with
\begin{eqnarray*}
&&c_{21}=\left(u\frac{\partial^{2} s}{\partial
u^{2}}+\frac{1}{2}\frac{\partial s}{\partial
u}\right)u_{x}^{2}+\left(u\frac{\partial
s}{\partial u}-\frac{1}{2}s\right)u_{xx}\\
&&c_{10}=\left(-u\frac{\partial^{2} s}{\partial u^{2}}-\frac{\partial
s}{\partial u}\right)u_{x}^{2}-u\frac{\partial s}{\partial u}u_{xx}\\
&&c_{11}=2u\frac{\partial s}{\partial u}u_{x}-su_{x}\\
&&c_{00}=-2u\frac{\partial s}{\partial u}u_{x}\\ 
&&c_{01}=0
\end{eqnarray*}

\noindent
Substituting the last equations in (\ref{silov}) we obtain
$(d_{2}X)_{xy}=0$.

\noindent
This means that all fields $X$ of degree 1 belong to $ker(d_{1}d_{2})$.

\noindent
Now we have to calculate the trivial field i.e. the fields
$X=d_{1}A+d_{2}B$.

\noindent
where 
\begin{eqnarray*}
&&A=\int_{S^{1}}A(u)dx\\
&&B=\int_{S^{1}}B(u)dx
\end{eqnarray*}

\noindent
Using the formulae
\begin{eqnarray}
d_{1}A=-\partial_{x}\frac{\delta A}{\delta u}
\end{eqnarray}

\noindent
and
\begin{eqnarray}
d_{2}B=-\partial_{x}\left(u\frac{\delta B}{\delta
u}\right)+\frac{1}{2}\frac{\delta B}{\delta u}u_{x}
\end{eqnarray}

\noindent
we get
\begin{eqnarray}
d_{1}A+d_{2}B=\left(-\frac{\partial^{2} A(u)}{\partial
u^{2}}-\frac{1}{2}\frac{\partial B(u)}{\partial u}-u\frac{\partial^{2}
B(u)}{\partial u^{2}}\right)u_{x}
\end{eqnarray}

\noindent
This shows immediately that all deformations are trivial.

\subsubsection{Explicit form of the deformations}

\noindent
We have to calculate $d_{1}X$ for an arbitary field of degree 1.

\noindent
$d_{2}X=0$ implies $d_{1}d_{2}X=-d_{2}d_{1}X=0$. Then there exists a 
vector field $\tilde{X}$ of degree 1 such that $d_{1}X=d_{2}\tilde{X}=0$.  

\noindent
This argument can be also used as an alternative proof of triviality of 
first order deformations. 

\subsection{Second order}

\noindent
Now we consider the deformation
$P_{2}=P^{(0)}_{2}-\lambda P^{(0)}_{1}+\epsilon^{2}P^{(2)}_{2}$ 
\footnote{The trivilaity of 
first order deformations means that 
we can always kill the term of first order in $\epsilon$ with a change 
of coordinates}.

\noindent
Also in this case the non trivial
deformations are elements of the group
$Ker(d_{1}d_{2})/\left(Im(d_{1})+Im(d_{2})\right)$.

\noindent
We start again considering the solutions of the equation
$d_{1}d_{2}X=0$.

\noindent
In this case the sum ($\ref{arnold}$) becomes
                                                            
\begin{eqnarray*}
&&(d_{2}X)_{xy}=\left(\frac{1}{2}\partial_{x}X+c_{10}+c_{21}+c_{32}\right)
\delta(x-y)+\left(X+c_{00}+c_{11}+c_{22}\right)\delta^{(1)}(x-y)+\\
&&+\left(c_{01}+c_{12}
\right)\delta^{(2)}(x-y)+c_{02}\delta^{(3)}(x-y)
\end{eqnarray*}

\noindent
where $X=s_{0}u_{xx}+s_{1}u_{x}^{2}$ and
\begin{eqnarray*}
&&c_{10}=-\partial_{x}\left(u\frac{\partial X}{\partial u}\right)\\
&&c_{21}=u\partial_{x}^{2}
\left(\frac{\partial X}{\partial u_{x}}\right)
-\frac{1}{2}\frac{\partial X}{\partial
u_{x}}u_{xx}+\frac{1}{2}\partial_{x}\left(\frac{\partial X}{\partial
u_{x}}\right)u_{x}\\
&&c_{32}=-u\partial_{x}^{3}\left(\frac{\partial X}{\partial
u_{xx}}\right)-\frac{1}{2}\frac{\partial X}{\partial
u_{xx}}u_{xxx}-\frac{1}{2}\partial_{x}^{2}\left(\frac{\partial X}{\partial
u_{xx}}\right)u_{x}\\
&&c_{00}=-2u\frac{\partial X}{\partial u}\\
&&c_{11}=2u\partial_{x}\left(\frac{\partial X}{\partial u_{x}}\right)
-\frac{\partial x}{\partial u_{x}}u_{x}\\
&&c_{22}=-3u\partial_{x}^{2}\left(\frac{\partial X}{\partial
u_{xx}}\right)-2\frac{\partial X}{\partial
u_{xx}}u_{xx}-\partial_{x}\left(\frac{\partial X}{\partial
u_{xx}}\right)u_{x}\\
&&c_{01}=0,\space c_{12}=-3u\partial_{x}\left(\frac{\partial X}{\partial
u_{xx}}\right)-3\frac{\partial X}{\partial u_{xx}}u_{x},\space c_{02}=
-2u\frac{\partial X}{\partial u_{xx}}
\end{eqnarray*}

\noindent
Hence we can write
\begin{equation}
(d_{2}X)_{xy}=\sum_{k=0}^{3}b_{k}\delta^{(k)}(x-y)
\end{equation}

\noindent
with $deg(b_{k})=3-k$. By straightforward calculation we get

\begin{eqnarray*}
&&b_{0}=au_{xxx}+bu_{x}u_{xx}+cu_{x}^{3}=\left(-2u\frac{\partial
s_{0}}{\partial u}+2us_{1}\right)u_{xxx}+\\
&&\left(s_{1}-4u\frac{\partial^{2} s_{0}}{\partial u^{2}}+4u\frac{\partial
s_{1}}{\partial u}-\frac{\partial s_{0}}{\partial u}\right)u_{x}u_{xx}+
\left(u\frac{\partial^{2} s_{1}}{\partial
u^{2}}+\frac{1}{2}\frac{\partial
s_{1}}{\partial u}-u\frac{\partial^{3} s_{0}}{\partial
u^{3}}-\frac{1}{2}\frac{\partial s_{0}}{\partial u^{2}}\right)u_{x}^{3}\\
&&b_{1}=du_{xx}+gu_{x}^{2}=\left(-5u\frac{\partial s_{0}}{\partial
u}+4us_{1}-s_{0}\right)u_{xx}+
\left(2u\frac{\partial s_{1}}{\partial u}-3u\frac{\partial^{2}
s_{0}}{\partial u^{2}}-\frac{\partial s_{0}}{\partial 
u}-s_{1}\right)u_{x}^{2}\\
&&b_{2}=hu_{x}=-3\left(u\frac{\partial s_{0}}{\partial
u}+s_{0}\right)u_{x}\\
&&b_{3}=l=-2us_{0}
\end{eqnarray*}

\noindent
Now we can calculate explicitly the equations $d_{1}d_{2}X=0$ in terms of
the coefficients $b_{0}, b_{1}, b_{2}, b_{3}$. By using the formula 
\begin{eqnarray*}
&&(d_{1}d_{2})_{xyz}=\f{1}{2}\sum_{t=0}^{3}\frac{\partial
\left(d_{2}X)_{xy}\right)}{\partial u^{t}(x)}\delta^{(t+1)}(x-z)
-\f{1}{2}\sum_{t=0}^{3}\frac{\partial
\left(d_{2}X)_{yx}\right)}{\partial u^{t}(y)}\delta^{(t+1)}(y-z)+\\       
&&\f{1}{2}\sum_{t=0}^{3}\frac{\partial 
\left(d_{2}X\right)_{zx}}{\partial
u^{t}(z)}\delta^{(t+1)}(z-y)-
\f{1}{2}\sum_{t=0}^{3}\frac{\partial \left(d_{2}X\right)_{xz}}{\partial
u^{t}(x)}\delta^{(t+1)}(x-y)+\\                                      
&&+\f{1}{2}\sum_{t=0}^{3}\frac{\partial
\left(d_{2}X\right)_{yz}}{\partial u^{t}(y)}\delta^{(t+1)}(y-x)
-\f{1}{2}\sum_{t=0}^{3}\frac{\partial
\left(d_{2}X\right)_{zy}}{\partial u^{t}(z)}\delta^{(t+1)}(z-x)
\end{eqnarray*}

\noindent
that is
\begin{eqnarray*}
&&(d_{1}d_{2})_{xyz}=\f{1}{2}\sum_{t=0}^{3}\left(\f{\d b_{0}}{\d 
u^{t}(x)} 
\right)\delta(x-y)\delta^{(t+1)}(x-z)+\f{1}{2}
\sum_{t=0}^{2}\left(\f{\d b_{1}}{\d u^{t}(x)}
\right)\delta'(x-y)\delta^{(t+1)}(x-z)+\\                               
&&+\f{1}{2}\sum_{t=0}^{1}\left(
\f{\d b_{2}}{\d u^{t}(x)}
\right)\delta(x-y)\delta^{(t+1)}(x-z)+                               
\f{1}{2}\f{\d b_{3}}{\d u(x)}
\delta^{(3)}(x-y)\delta'(x-z)+\\   
&&-\f{1}{2}\sum_{t=0}^{3}\left(\f{\d b_{0}}{\d
u^{t}(y)}
\right)\delta(y-x)\delta^{(t+1)}(y-z)-
\f{1}{2}\sum_{t=0}^{2}\left(\f{\d b_{1}}{\d u^{t}(y)}
\right)\delta'(y-x)\delta^{(t+1)}(y-z)-\\
&&\f{1}{2}-\sum_{t=0}^{1}\left(
\f{\d b_{2}}{\d u^{t}(y)}
\right)\delta(y-x)\delta^{(t+1)}(y-z)-
\f{1}{2}\f{\d b_{3}}{\d u(y)}
\delta^{(3)}(y-x)\delta'(y-z)+...                           
\end{eqnarray*}

\noindent
In order to obtain a sum where appear only terms with 
$\delta^{(i)}(x-y)\delta^{(j)}(x-z)$ and where the coefficients depend 
only on $x$ we use the identity 

\begin{eqnarray}
\delta(x-y)\delta(x-z)=\delta(y-x)\delta(y-z)=\delta(z-x)\delta(z-y)
\end{eqnarray}

\noindent
and formula (\ref{formula}). For example we can write
\begin{eqnarray*}
&&-\f{1}{2}\sum_{t=0}^{3}\left(\f{\d b_{0}}{\d
u^{t}(y)}
\right)\delta(y-x)\delta^{(t+1)}(y-z)=\\
&&=-(-1)^{t+1}\f{1}{2}\sum_{t=0}^{3}\left(\f{\d b_{0}}{\d
u^{t}(y)}
\right)\d^{t+1}_{z}\left(\delta(y-x)\delta(y-z)\right)=\\
&&-(-1)^{t+1}\f{1}{2}\sum_{t=0}^{3}\left(\f{\d b_{0}}{\d
u^{t}(y)}
\right)\d^{t+1}_{z}\left(\delta(x-y)\delta(x-z)\right)=\\  
&&=-\f{1}{2}\sum_{t=0}^{3}\left(\f{\d b_{0}}{\d
u^{t}(y)}
\right)\delta(x-y)\delta^{(t+1)}(x-z)=\\
&&-\f{1}{2}\sum_{t=0}^{3}\left(\f{\d b_{0}}{\d
u^{t}(x)}
\right)\delta(x-y)\delta^{(t+1)}(x-z)                                         
\end{eqnarray*}

\noindent      
In this way, after a long but elementary calculation 
we obtain 
\begin{equation}
d_{1}d_{2}X=2f\delta\wedge\delta^{3}+3f\delta^{1}\wedge\delta^{2}
+3f_{x}\delta\wedge\delta^{2}+f_{xx}\delta\wedge\delta^{1}
\end{equation}

\noindent
where 
$\delta^{i}\wedge\delta^{j}
=\f{1}{2}\left(\delta^{(i)}(x-y)\delta^{(j)}(x-z)
-\delta^{(j)}(x-y)\delta^{(i)}(x-z)\right)$ 
and

\begin{equation}
f=\f{\d b_{1}}{\d u_{x}}-\d_{x}\left(\f{\d b_{1}}{\d 
u_{xx}}\right)-\f{\d b_{2}}{\d u}+\d_{x}\left(\f{\d b_{3}}{\d u}\right)
\end{equation}
 
\noindent
Therefore the equation $d_{1}d_{2}X=0$ is equivalent to the equation 
$f=0$. By substituting $b_{1}$, $b_{2}$, $b_{3}$ in this equation we get 
the condition
\begin{equation}
s_{1}=\f{\d s_{0}}{\d u}
\end{equation}

\noindent
\subsubsection{Trivial deformations}

\noindent
The differentials $d_{1}$ and $d_{2}$ increase the degree by one. 
Because the 
degree of $X$ is 2 the trivial deformations can be written as
 $X=d_{1}A+d_{2}B=0$ where the degree of the densities of $A$ and $B$ is 
1. But the variational derivative of local functionals with densities of 
degree 1 vanishes:
\begin{equation*}
\f{\delta \int_{S^{1}}(f(u)u_{x})dx}{\delta 
u}=\sum_{i=0}(-1)^{i}\d_{x}^{i}\left(\f{\d 
f}{\d u^{(i)}}\right)=
\f{\d f}{\d u}u_{x}-\d_{x}(f(u))=0
\end{equation*}

\noindent
Then all the deformations $d_{1}X^{2}_{2}$ (with
$X_{2}^{2}=s_{0}u_{xx}+\frac{\partial s_{0}}{\partial
u}u_{x}^{2}$) are not trivial.

\subsubsection{Deformations: explicit form}

\noindent
By using the formula (\ref{formula2}), we get
\begin{eqnarray*}
&&d_{1}X=\sum_{s}\left(\left(\d^{s}_{y}\delta'(y-x)\right)\f{\d}{\d 
u^{(s)}(y)}(su_{yy}+\f{\d 
s}{\d u}u_{y}^{2})-\left(\d^{s}_{x}\delta'(x-y)\right)\f{\d}{\d
u^{(s)}(x)}(su_{xx}+\f{\d s}{\d u}u_{x}^{2})\right)=\\
&&\left(\f{\d s}{\d u}u_{yy}+\f{\d^{2} s}{\d 
u^{2}}u_{y}^{2}\right)\delta^{(1)}(y-x)
-\left(\f{\d s}{\d u}u_{xx}+\f{\d^{2} s}{\d u^{2}}u_{x}^{2}
\right)\delta^{(1)}(x-y)+\\
&&\left(2\f{\d s}{\d u}u_{y}\right)\delta^{(2)}(y-x)
-\left(2\f{\d s}{\d u}u_{x}\right)\delta^{(2)}(x-y)+s(y)\delta^{(3)}(y-x)
-s(x)\delta^{(3)}(x-y)
\end{eqnarray*}

\noindent
By observing that 
\begin{equation*}
\f{\d s}{\d u}u_{xx}+\f{\d^{2} s}{\d u^{2}}u_{x}^{2}=\d_{x}^{2}s
\end{equation*}  

\noindent
and by using the identity (\ref{formula}) it is easy to get the formula

\begin{equation}
P^{(2)}_{2}=d_{1}X^{(2)}_{2}=-2s\delta^{3}(x-y)-3\d_{x}s\delta^{2}(x-y)
-\d_{x}^{2}s\delta^{1}(x-y)
\end{equation}                                                                                                            

\subsection{Third order}

\noindent
The condition of compatibility $[P_{1},P_{2}]=0$ implies that
$P^{3}_{2}=d_{1}X^{(3)}_{2}$ and the Jacoby identity
$[P_{2},P_{2}]=0$
implies that $d_{2}P^{(3)}_{2}=-d_{1}d_{2}X^{(3)}_{2}=0$.

\noindent
{\bf Remark}

\noindent
There are no conditions containing the second order deformations.\\

\noindent
We start again  calculating $d_{2}X$. From (\ref{formula2}) it follows:
\begin{eqnarray*}
&&\left(d_{2}X\right)_{xy}=\\
&&\left(\f{1}{2}\d_{x}X+c_{10}+c_{21}+c_{32}
+c_{43}\right)
\delta(x-y)+\left(X+c_{00}+c_{11}+c_{22}+c_{33}\right)
\delta^{(1)}(x-y)+\\
&&+\left(c_{01}+c_{12}+c_{23}\right)\delta^{(2)}(x-y)+\left(c_{02}
+c_{13}\right)\delta^{(3)}(x-y)+c_{03}\delta^{(4)}(x-y)
\end{eqnarray*}

\noindent
where $X=s_{0}u_{xxx}+s_{1}u_{x}u_{xx}+s_{2}u_{x}^{3}$ and
\begin{eqnarray*}
&&c_{10}=-\d_{x}\left(u\f{\d X}{\d u}\right)\\
&&c_{21}=u\d_{x}^{2}\left(\f{\d X}{\d u_{x}}\right)-\f{1}{2}
\f{\d X}{\d
u_{x}}u_{xx}+\f{1}{2}u_{x}\d_{x}\left(\f{\d X}{\d u_{x}}
\right)\\
&&c_{32}=-u\d_{x}^{3}\left(\f{\d X}{\d
u_{xx}}\right)-\f{1}{2}\f{\d X}{\d
u_{xx}}u_{xxx}-\f{1}{2}u_{x}\d_{x}^{2}\left(\f{\d X}{\d
u_{xx}}\right)\\
&&c_{43}=u\d_{x}^{4}\left(\f{\d X}{\d
u_{xxx}}\right)-\f{1}{2}\f{\d X}{\d
u_{xxx}}u_{xxxx}+\f{1}{2}u_{x}\d_{x}^{3}\left(\f{\d X}{\d
u_{xxx}}\right)\\
&&c_{00}=-2u\f{\d X}{\d u}\\
&&c_{11}=2u\d_{x}\left(\f{\d X}{\d u_{x}}\right)-\f{\d X}{\d
u_{x}}u_{x}\\
&&c_{22}=-3u\d_{x}^{2}\left(\f{\d X}{\d u_{xx}}\right)-2\f{\d
X}{\d
 u_{xx}}u_{xx}-u_{x}\d_{x}\left(\f{\d X}{\d u_{xx}}\right)\\
&&c_{33}=4u\d_{x}^{3}\left(\f{\d X}{\d
u_{xxx}}\right)-\f{5}{2}\f{\d
X}{\d
u_{xxx}}u_{xxx}+\f{3}{2}u_{x}\d_{x}^{2}\left(\f{\d X}{\d
u_{xxx}}\right)\\
&&c_{01}=0\\
&&c_{12}-3u\d_{x}\left(\f{\d X}{\d u_{xx}}\right)-3\f{\d X}{\d
u_{xx}}u_{x}\\
&&c_{23}=6u\d_{x}^{2}\left(\f{\d X}{\d
u_{xxx}}\right)-\f{9}{2}\f{\d X}{\d
u_{xxx}}u_{xx}+\f{3}{2}u_{x}\d_{x}\left(\f{\d X}{\d
u_{xxx}}\right)\\
&&c_{13}=4u\d_{x}\left(\f{\d X}{\d u_{xxx}}\right)-3\f{\d X}{\d
u_{xxx}}u_{x}\\                                                
&&c_{03}=0
\end{eqnarray*}

\noindent
Then we can write
\begin{equation}
\left(d_{2}X\right)_{xy}=\sum_{k=0}^{3}b_{k}\delta^{(k)}(x-y)
\end{equation}

\noindent
with $deg(b_{k})=4-k$ and
\begin{eqnarray*}
&&b_{0}=au_{xxxx}+bu_{x}u_{xxx}+c\left(u_{xx}\right)^{2}+du_{x}^{2}
u_{xx}+gu_{x}^{4}=\\
&&=0+\left(-3u\f{\d s_{1}}{\d u}+3u\f{\d^{2} s_{0}}{\d
u^{2}}+6us_{2}\right)\left(u_{x}u_{xxx}
+\left(u_{xx}\right)^{2}\right)+\\
&&\left(-6u\f{\d^{2} s_{1}}{\d u^{2}}+12u\f{\d s_{2}}{\d
u}-\f{3}{2}\f{\d s_{1}}{\d u}+3s_{2}+6u\f{\d^{3} s_{0}}{\d
u^{3}}+\f{3}{2}\f{\d^{2} s_{0}}{\d
u^{2}}\right)u_{x}^{2}u_{xx}+\\
&&+\left(2u\f{\d^{2} s_{2}}{\d u^{2}}+\f{\d \w}{\d
u}-\f{1}{2}\f{\d^{2} \t}{\d u^{2}}+u\f{\d^{4} \s}{\d
u^{4}}+\f{1}{2}\f{\d^{3}\s}{\d u^{3}}-u\f{\d^{3}\t}{\d
u^{3}}\right)u_{x}^{4}\\
&&b_{1}=hu_{xxx}+lu_{x}u_{xx}+mu_{x}^{3}=\\
&&=\left(2u\f{\d\s}{\d
u}-u\t-\f{3}{2}\s\right)u_{xxx}+\left(12u\w-3\t-9u\f{\d\t}{\d
u}+12u\f{\d^{2}\s}{\d u^{2}}+\f{3}{2}\f{\d\s}{\d u}\right)u_{x}u_{xx}+\\
&&+\left(4u\f{\d\w}{\d u}-2\w-3u\f{\d^{2}\t}{\d u^{2}}-\f{\d\t}{\d
u}+4u\f{\d^{3}\s}{\d u^{3}}+\f{3}{2}\f{\d^{2}\s}{\d
u^{2}}\right)u_{x}^{3}\\
&&b_{2}=pu_{xx}+qu_{x}^{2}=\\
&&=\left(-3u\t+6u\f{\d\s}{\d u}-\f{9}{2}\s\right)u_{xx}+
\left(-3u\f{\d\t}{\d u}-3\t+6u\f{\d^{2}\s}{\d u^{2}}+\f{3}{2}\f{\d\s}{\d
u}\right)u_{x}^{2}\\
&&b_{3}=ku_{x}=\left(-2u\t+4u\f{\d\s}{\d u}-3\s\right)u_{x}
\end{eqnarray*}                                                      

\subsubsection{Deformations}

\noindent
In terms of coefficients $b_{k}$ the equation $(d_{1}d_{2}X)_{xy}=0$ can
be written simply by adding to the equation obtained in the previous 
case the terms
\begin{eqnarray*}
&&\f{1}{2}\left(\f{\d b_{0}}{\d
u_{xxxx}}\right)\delta(x-y)\delta^{(5)}(x-z)+\f{1}{2}\left(\f{\d
b_{1}}{\d u_{xxx}}\right)\delta^{(1)}(x-y)\delta^{(4)}(x-z)\\
&&+\f{1}{2}\left(\f{\d b_{2}}{\d
u_{xx}}\right)\delta^{(2)}(x-y)\delta^{(3)}(x-y)
+\f{1}{2}\left(\f{\d b_{3}}{\d 
u_{x}}\right)\delta^{(3)}(x-y)\delta^{(2)}(x-z)\\
&&-\f{1}{2}\left(\f{\d b_{0}}{\d
u_{yyyy}}\right)\delta(y-x)\delta^{(5)}(y-z)-\f{1}{2}\left(\f{\d
b_{1}}{\d u_{yyy}}\right)\delta^{(1)}(y-x)\delta^{(4)}(y-z)\\
&&-\f{1}{2}\left(\f{\d b_{2}}{\d
u_{yy}}\right)\delta^{(2)}(y-x)\delta^{(3)}(y-z)
-\f{1}{2}\left(\f{\d b_{3}}{\d 
u_{y}}\right)\delta^{(3)}(y-x)\delta^{(2)}(y-z)\\
&&+\f{1}{2}\left(\f{\d b_{0}}{\d
u_{zzzz}}\right)\delta(z-x)\delta^{(5)}(z-y)+\f{1}{2}\left(\f{\d
b_{1}}{\d u_{zzz}}\right)\delta^{(1)}(z-x)\delta^{(4)}(z-y)\\
&&+\f{1}{2}\left(\f{\d b_{2}}{\d
u_{zz}}\right)\delta^{(2)}(z-x)\delta^{(3)}(z-y)
+\f{1}{2}\left(\f{\d b_{3}}{\d 
u_{z}}\right)\delta^{(3)}(z-x)\delta^{(2)}(z-y)\\ 
&&-\f{1}{2}\left(\f{\d b_{0}}{\d
u_{xxxx}}\right)\delta(x-z)\delta^{(5)}(x-y)-\f{1}{2}\left(\f{\d
b_{1}}{\d u_{xxx}}\right)\delta^{(1)}(x-z)\delta^{(4)}(x-y)\\
&&-\f{1}{2}\left(\f{\d b_{2}}{\d
u_{xx}}\right)\delta^{(2)}(x-z)\delta^{(3)}(x-y)
-\f{1}{2}\left(\f{\d b_{3}}{\d 
u_{x}}\right)\delta^{(3)}(x-z)\delta^{(2)}(x-y)\\ 
&&+\f{1}{2}\left(\f{\d b_{0}}{\d
u_{yyyy}}\right)\delta(y-z)\delta^{(5)}(y-x)+\f{1}{2}\left(\f{\d
b_{1}}{\d u_{yyy}}\right)\delta^{(1)}(y-z)\delta^{(4)}(y-x)\\
&&+\f{1}{2}\left(\f{\d b_{2}}{\d
u_{yy}}\right)\delta^{(2)}(y-z)\delta^{(3)}(y-x)
+\f{1}{2}\left(\f{\d b_{3}}{\d 
u_{y}}\right)\delta^{(3)}(y-z)\delta^{(2)}(y-x)\\ 
&&-\f{1}{2}\left(\f{\d b_{0}}{\d
u_{zzzz}}\right)\delta(z-y)\delta^{(5)}(z-x)-\f{1}{2}\left(\f{\d
b_{1}}{\d u_{zzz}}\right)\delta^{(1)}(z-y)\delta^{(4)}(z-x)\\
&&-\f{1}{2}\left(\f{\d b_{2}}{\d
u_{zz}}\right)\delta^{(2)}(z-y)\delta^{(3)}(z-x)
+\f{1}{2}\left(\f{\d b_{3}}{\d 
u_{z}}\right)\delta^{(3)}(z-y)\delta^{(2)}(z-x) 
\end{eqnarray*}

\noindent
By introducing the function $f$ and $g$ defined by the formulae
\begin{eqnarray*}
&&f=\f{\d b_{1}}{\d u_{x}}-\d_{x}\left(\f{\d b_{1}}{\d 
u_{xx}}\right)+\d_{x}^{2}\left(\f{\d b_{1}}{\d u_{xxx}}\right)-\f{\d 
b_{2}}{\d u}+\d_{x}\left(\f{\d b_{3}}{\d u}\right)\\
&&g=\f{\d b_{1}}{\d u_{xxx}}-\f{\d b_{2}}{\d u_{xx}}+\f{\d b_{3}}{\d u_{x}}
\end{eqnarray*}

\noindent
we can write the result in the form
\begin{equation}
d_{1}d_{2}X=\sum f_{ij}\delta^{i}\wedge\delta^{j}
\end{equation}

\noindent
where
\begin{eqnarray*}
&&f_{05}=f_{23}=2g\\
&&f_{14}=5g\\
&&f_{04}=5g_{x}\\
&&f_{13}=8g_{x}\\
&&f_{03}=2f+4g_{xx}\\
&&f_{12}=3f+3g_{xx}\\ 
&&f_{02}=3f_{x}+g_{xxx}\\
&&f_{01}=f_{xx}
\end{eqnarray*}

\noindent
Therefore the equation $d_{1}d_{2}X=0$ is equivalent to the equations 
$f=0$ and $g=0$.

\noindent
By substituting the coefficients $b_{1}$, $b_{2}$ and $b_{3}$ in these 
equations we obtain that the last equation is identically satisfied and 
the first is equivalent to the condition
\begin{equation}
2s_{2}-\f{\d s_{1}}{\d u}+\f{\d^{2} s_{0}}{\d u^{2}}
\label{equation1}                                                              
\end{equation}

\noindent
\subsubsection{Trivial deformations}

\noindent
In this case trivial deformations are $d_{1}A+d_{2}B$ with
\begin{eqnarray*}
&&A=\int_{S^{1}}(A_{0}(u)u_{xx}+A_{1}(u)u_{x}^{2})dx\\
&&B=\int_{S^{1}}(B_{0}(u)u_{xx}+B_{1}(u)u_{x}^{2})dx
\end{eqnarray*}

\noindent
By using formula (\ref{formula1}) we get 
\begin{eqnarray*}
&&d_{1}A+d_{2}B=\left(-2\left(\f{\d A_{0}}{\d
u}-A_{1}\right)-2u\left(\f{\d B_{0}}{\d
u}-B_{1}\right)\right)u_{xxx}+\\
&&\left(-4\left(\f{\d^{2} A_{0}}{\d u^{2}}-\f{\d A_{1}}{\d
u}\right)-4u\left(\f{\d^{2}B_{0}}{\d u^{2}}-\f{\d B_{1}}{\d
u}\right)-\left(\f{\d B_{0}}{\d u}-B_{1}\right)\right)+\\
&&+\left(-\left(\f{\d^{3} A_{0}}{\d u^{3}}-\f{\d^{2} A_{1}}
{\d u^{2}}\right)-u\left(\f{\d^{3}B_{0}}{\d u^{3}}-\f{\d^{2}
B_{1}}{\d u^{2}}\right)-\f{1}{2}\left(\f{\d^{2}B_{0}}{\d
u^{2}}-\f{\d
B_{1}}{\d u}\right)\right)
\end{eqnarray*}

\noindent
If we call $\tilde{A}:=A_{1}-\f{\d A_{0}}{\d u}$ and
$\tilde{B}:=B_{1}-\f{\d B_{0}}{\d u}$, we can write
\begin{eqnarray*}
&&d_{1}A+d_{1}B=\\
&&\left(2\tilde{A}+2u\tilde{B}\right)
u_{xxx}+
\left(4\f{\d \tilde{A}}{\d u}+4u\f{\d \tilde{B}}{\d
u}+\tilde{B}\right)
u_{x}u_{xx}
+\left(\f{\d^{2}\tilde{A}}{\d u^{2}}+4u\f{\d^{2}
\tilde{B}}{\d u^{2}}+\f{\d\tilde{B}}{\d u}\right)
u_{x}^{3}
\end{eqnarray*}                                                             

\noindent
Now we can prove that all deformations $P^{(3)}_{2}$ are trivial.

\noindent
{\bf Proof}

\noindent
The trivial deformations satisfy equation (\ref{equation1}). In fact
\begin{eqnarray*}
2\left(\f{\d^{2}\tilde{A}}{\d u^{2}}+4u\f{\d^{2}
\tilde{B}}{\d u^{2}}+\f{\d\tilde{B}}{\d u}\right)-\f{\d}{\d u}
\left(4\f{\d \tilde{A}}{\d u}+4u\f{\d \tilde{B}}{\d
u}+\tilde{B}\right)+\f{\d^{2}}{\d u^{2}}
\left(2\tilde{A}+2u\tilde{B}\right)=0
\end{eqnarray*}

\noindent
Every field $X=s_{0}u_{xxx}+s_{1}u_{x}u_{xx}+s_{2}u_{x}^{3}$
with coefficients $s_{0}$, $s_{1}$, $s_{2}$ satisfying
equation (\ref{equation1}) can be written as $d_{1}A+d_{2}B$ by
choosing
\begin{eqnarray*}
&&\tilde{A}=\f{s}{2}-\f{u}{3}\left(2\f{\d s_{0}}{\d u}
-s_{1}\right)\\
&&\tilde{B}=\f{1}{3}\left((2\f{\d s_{0}}{\d u}-s_{1}\right)
\end{eqnarray*}

\noindent
and this choice is always possible.

\subsubsection{Deformations: explicit form}

\noindent
By using the formula (\ref{formula2})
\begin{eqnarray*}
&&d_{1}X=
\sum_{s\geq 0}\left(\left(\d^{s}_{y}\delta'(y-x)\right)\f{\d}{\d
u^{(s)}(y)}(s_{0}u_{yyy}+s_{1}u_{y}u_{yy}+s_{2}u_{y}^{3})+\right.\\
&&\left.-\left(\d^{s}_{x}\delta'(x-y)\right)\f{\d}{\d
u^{(s)}(x)}(s_{0}u_{xxx}+s_{1}u_{x}u_{xx}+s_{2}u_{x}^{3})\right)
\end{eqnarray*}
                                                                                
\noindent
and condition (\ref{equation1}) it is easy to get the formula

\begin{eqnarray*}
&&P^{(3)}_{2}=d_{1}X_{2}^{(3)}=
-2t\delta^{(3)}(x-y)-3\d_{x}t\delta^{(2)}(x-y)
-\d_{x}^{2}t\delta^{(1)}(x-y)
\end{eqnarray*}                                                    

\noindent
where $t$ is an arbitrary differential polynomial of degree 1.

\noindent
\subsection{Fourth
order}

\noindent
In this section we consider the case:
$P_{2}=P_{2}^{(0)}+\epsilon^{2}P^{(2)}_{2}+\epsilon^{4}
P^{(4)}_{2}$.\footnote{The third order term can be always killed
by changing coordinates and the second order term is                 

$P_{2}^{(2)}=d_{1}X_{2}^{(2)}$ with
$X^{(2)}_{2}=su_{xx}+\f{\d s}{\d u}u_{x}^{2}$}

\noindent
The compatibility condition implies
$P_{2}^{(4)}=d_{1}X_{2}^{(4)}$ and the Jacoby identity 
$[P_{2},P_{2}]=o(\epsilon^{4})$ implies
\begin{equation*}
d_{1}d_{2}X^{(4)}_{2}-\f{1}{2}
\left[d_{1}X_{2}^{(2)},
d_{1}X_{2}^{(2)}\right]=0
\end{equation*}

\noindent
\subsubsection{Deformation}

\noindent
First of all we consider the term $\left[d_{1}X_{2}^{(2)},
d_{1}X_{2}^{(2)}\right]$.
By using the formula (\ref{formula3}) we get
\begin{eqnarray}
&&\nonumber \left[d_{1}X_{2}^{(2)},
d_{1}X_{2}^{(2)}\right]=\\
&&\nonumber =\f{\d \left(d_{1}X_{2}^{(2)}\right)_{xy}}{\d
u^{s}(x)}\d_{x}^{s}\left(d_{1}X_{2}^{(2)}\right)_{xz}
-\f{\d \left(d_{1}X_{2}^{(2)}\right)_{yx}}{\d
u^{s}(y)}\d_{y}^{s}\left(d_{1}X_{2}^{(2)}\right)_{yz}+\\                                                          
&&\nonumber +\f{\d \left(d_{1}X_{2}^{(2)}\right)_{zx}}{\d
u^{s}(z)}\d_{z}^{s}\left(d_{1}X_{2}^{(2)}\right)_{zy}                                                          
-\f{\d \left(d_{1}X_{2}^{(2)}\right)_{xz}}{\d
u^{s}(x)}\d_{x}^{s}\left(d_{1}X_{2}^{(2)}\right)_{xy}+\\
&& +\f{\d \left(d_{1}X_{2}^{(2)}\right)_{yz}}{\d
u^{s}(y)}\d_{y}^{s}\left(d_{1}X_{2}^{(2)}\right)_{yx}
-\f{\d \left(d_{1}X_{2}^{(2)}\right)_{zy}}{\d
u^{s}(z)}\d_{z}^{s}\left(d_{1}X_{2}^{(2)}\right)_{zx}    
\label{equation2}
\end{eqnarray}

\noindent
Let us focus our attention on the first term. By straightforward 
calculation we get
\begin{equation}
\f{\d \left(d_{1}X_{2}^{(2)}\right)_{xy}}{\d
u^{s}(x)}\d_{x}^{s}\left(d_{1}X_{2}^{(2)}\right)_{xz}=
\sum_{i=1,...,3;j=1,...,6-i} b_{ij}\delta^{(i)}(x-y)\delta^{(j)}(x-z)
\end{equation}

\noindent
where
\begin{eqnarray*}
&&b_{11}=\left(6\f{\d^{3} s}{\d u^{3}}\f{\d^{2} s}{\d u^{2}}+2\f{\d s}
{\d u}\f{\d^{4} s}{\d u^{4}}\right)u_{x}^{4}+\left(14\left(\f{\d^{2} s}{\d 
u^{2}}\right)^{2}+14\f{\d^{3} s}{\d u^{3}}\f{\d s}{\d 
u}\right)u_{x}^{2}u_{xx}+\\
&&\left(8\f{\d^{2} s}{\d u^{2}}\f{\d s}{\d 
u}\right)u_{xx}^{2}+\left(12\f{\d^{2} s}{\d u^{2}}\f{\d  s}{\d 
u}\right)u_{x}u_{xxx}+2\left(\f{\d s}{\d u}\right)^{2}u_{xxxx}\\
&&b_{12}=\left(16\f{\d^{3} s}{\d u^{3}}\f{\d s}{\d u}+16\left(\f{\d^{2} 
s}{\d u^{2}}\right)^{2}\right)u_{x}^{3}+\left(52\f{\d^{2} s}{\d 
u^{2}}\f{\d s}{\d u}\right)u_{x}u_{xx}+10\left(\f{\d s}{\d 
u}\right)^{2}u_{xxx}\\
&&b_{21}=\left(6\f{\d^{3} s}{\d u^{3}}\f{\d s}{\d u}+6\left(\f{\d^{2}
s}{\d u^{2}}\right)^{2}\right)u_{x}^{3}+\left(24\f{\d^{2} s}{\d
u^{2}}\f{\d s}{\d u}\right)u_{x}u_{xx}+6\left(\f{\d s}{\d
u}\right)^{2}u_{xxx}\\                                                   
&&b_{13}=\left(4\f{\d^{3} s}{\d u^{3}}s+38\f{\d^{2} s}{u^{2}}\f{\d s}{\d 
u}\right)u_{x}^{2}+\left(4\f{\d^{2} s}{\d u^{2}}s+18\left(\f{\d s}{\d 
u}\right)^{2}\right)u_{xx}\\
&&b_{31}=\left(4\f{\d^{2} s}{u^{2}}\f{\d s}{\d u}\right)u_{x}^{2}
+4\left(\f{\d s}{\d u}\right)^{2}u_{xx}\\                                           
&&b_{22}=\left(42\f{\d^{2} s}{u^{2}}\f{\d s}{\d u}\right)u_{x}^{2}
+24\left(\f{\d s}{\d u}\right)^{2}u_{xx}\\                           
&&b_{23}=\left(12\f{\d^{2} s}{\d u^{2}}s\right)u_{x}+30\left(\f{\d s}{\d 
u}\right)^{2}\\
&&b_{32}=12\left(\f{\d s}{\d u}\right)^{2}u_{x}\\
&&b_{14}=\left(8\f{\d^{2} s}{\d u^{2}}s\right)u_{x}+14\left(\f{\d s}{\d
u}\right)^{2}\\
&&b_{24}=12\f{\d s}{\d u}s\\
&&b_{33}=8\f{\d s}{\d u}s\\
&&b_{15}=4\f{\d s}{\d u}s                                                         
\end{eqnarray*}             

\noindent
The other terms in (\ref{equation2}) have the same form. The only 
difference 
is that the variables $x$, $y$ and $z$ play a different role. 
Therefore we can apply the usual tricks and write an expression 
containing only terms with $\delta^{(i)}(x-y)\delta^{(j)}(x-z)$.
For example we can write

\begin{eqnarray*}
&&-\f{\d \left(d_{1}X_{2}^{(2)}\right)_{yx}}{\d
u^{s}(y)}\d_{y}^{s}\left(d_{1}X_{2}^{(2)}\right)_{yz}=
-\sum_{ij}b_{ij}(y)\delta^{(i)}(x-y)\delta^{(j)}(x-z)=\\
&&-(-1)^{i+j}\sum_{ij}b_{ij}(y)\d_{x}^{i}\d_{z}^{j}\left(\delta(y-x)
\delta(y-z)\right)=\\
&&=-(-1)^{i+j}\sum_{ij}b_{ij}(y)\d_{x}^{i}\d_{z}^{j}\left(\delta(x-y)
\delta(x-z)\right)=\\
&&(-1)^{i+1}\sum_{ij}b_{ij}(y)\d_{x}^{i}\left(\delta(x-y)
\delta^{(j)}(x-z)\right)=\\
&&=(-1)^{i+1}\sum_{ij}b_{ij}(y)\sum_{k=0}^{i}\bin{i}{k}\delta^{(k)}(x-y)
\delta^{(j+i-k)}(x-z)=\\
&&(-1)^{i+1}\sum_{ij}\sum_{k=0}^{i}\bin{i}{k}\sum_{l=0}^{k}
\bin{k}{l}b_{ij}(x)^{(l)}\delta^{(k-l)}(x-y)
\delta^{(j+i-k)} (x-z)                                             
\end{eqnarray*}

\noindent
The final result is

\begin{eqnarray*}
&&\left[d_{1}X_{2}^{(2)},d_{1}X_{2}^{(2)}\right]=16\left(s\f{\d
s}{\d u}\right)\delta^{1}\wedge\delta^{5}+40\left(s\f{\d s}{\d
u}\right)\delta^{2}\wedge\delta^{4}+40\d_{x}\left(s\f{\d s}{\d
u}\right)\delta^{1}\wedge\delta^{4}+\\
&&+48\d_{x}\left(s\f{\d s}{\d
u}\right)\delta^{2}\wedge\delta^{3}+32\d^{2}_{x}\left(s\f{\d
s}{\d u}\right)\delta^{1}\wedge\delta^{3}+8\d_{x}^{3}\left(s\f{\d
s}{\d u}\right)\delta^{1}\wedge\delta^{2}
\end{eqnarray*}                                                      

\noindent
The calculation of the term $d_{2}X^{(4)}_{2}$ can be done as above. In 
this case we have
\begin{eqnarray*}
&&\left(d_{2}X^{(4)}_{2}\right)_{xy}=\left(\f{1}{2}\d_{x}X
+c_{10}+c_{21}+c_{32}+c_{43}+c_{54}\right)\delta(x-y)\\
&&\left(X+c_{00}+c_{11}+c_{22}+c_{33}+c_{44}\right)\delta^{(1)}(x-y)
+\left(c_{01}+c_{12}+c_{23}+c_{34}\right)\delta^{(2)}(x-y)\\
&&+\left(c_{02}+c_{13}+c_{24}\right)\delta^{(3)}(x-y)
+\left(c_{03}+c_{14}\right)\delta^{(4)}(x-y)+c_{04}\delta^{(5)}(x-y)
\end{eqnarray*}

\noindent
where
$X^{(4)}_{2}=X=s_{0}\uxxxx+s_{1}\ux\uxxx
+s_{2}(\uxx)^{2}+s_{3}\ux^{2}\uxx+s_{4}\ux^{4}$ and the
coefficients $c_{ij}$ have the same expression in terms of
$X^{(4)}_{2}$ that in the previous case, except for the new
coefficients
\begin{eqnarray*}
&&c_{54}=-u\d^{5}_{x}\left(\f{\d X}{\d
\uxxxx}\right)-\f{1}{2}\f{\d X}{\d
\uxxxx}\uxxxxx-\f{1}{2}\ux\d_{x}^{4}\left(\f{\d X}{\d
\uxxxx}\right)\\
&&c_{44}=-5u\d^{4}_{x}\left(\f{\d X}{\d
\uxxxx}\right)-3\f{\d X}{\d
\uxxxx}\uxxxx-2\ux\d_{x}^{3}\left(\f{\d X}{\d
\uxxxx}\right)\\
&&c_{34}=-10u\d^{3}_{x}\left(\f{\d X}{\d
\uxxxx}\right)-7\f{\d X}{\d
\uxxxx}\uxxx-3\ux\d_{x}^{2}\left(\f{\d X}{\d
\uxxxx}\right)\\
&&c_{24}=-10u\d^{2}_{x}\left(\f{\d X}{\d
\uxxxx}\right)-8\f{\d X}{\d
\uxxxx}\uxx-2\ux\d_{x}\left(\f{\d X}{\d
\uxxxx}\right)\\                                                             
&&c_{14}=-5u\d_{x}\left(\f{\d X}{\d
\uxxxx}\right)-5\f{\d X}{\d
\uxxxx}\ux\\
&&c_{04}=-2u\f{\d X}{\d\uxxxx}
\end{eqnarray*}

\noindent
Consequently we can write                                                  

\begin{equation*}
\left(d_{2}X^{(4)}_{2}\right)_{xy}=\sum_{k=0}^{5}b_{k}\delta^{(k)}(x-y)
\end{equation*}

\noindent
with $deg(b_{k})=5-k$ and
\begin{eqnarray*}
&&b_{0}=a_{0}\uxxxxx+a_{1}\ux\uxxxx+a_{2}\uxx\uxxx+a_{3}\ux^{2}\uxxx
+a_{4}\ux(\uxx)^{2}+a_{5}\ux^{3}\uxx+a_{6}\ux^{5}=\\
&&\left(2us_{1}-2us_{2}-2u\f{\d s_{0}}{\d
u}\right)\uxxxxx+\left(s_{1}-\f{\d s_{0}}{\d u}-s_{2}
-6u\f{\d s_{2}}{\d u}+6u\f{\d
s_{1}}{\d
u}-6u\f{\d^{2} s_{0}}{\d u^{2}}\right)\ux\uxxxx\\
&&+\left(-10u\f{\d s_{2}}{\d u}+10u\f{\d s_{1}}{\d
u}-10u\f{\d^{2} s_{0}}{\d u^{2}}\right)\uxx\uxxx+\left(2\f{\d
s_{1}}{\d u}-2\f{\d s_{2}}{\d u}-2\f{\d^{2}s_{0}}{\d
u^{2}}
+12us_{4}
-4u\f{\d s_{3}}{\d u}+\right.\\
&&\left.-6u\f{\d^{2}s_{2}}{\d u^{2}}+10u\f{\d^{2}s_{1}}{\d u^{2}}
-10u\f{\d^{3}s_{0}}{\d
u^{3}}\right)\ux^{2}\uxxx
+\left(-\f{3}{2}\f{\d s_{2}}{\d u}+
\f{3}{2}\f{\d s_{1}}{\d u}-\f{3}{2}\f{\d^{2} s_{0}}{\d
u^{2}}+24us_{4}
-8u\f{\d s_{3}}{\d u}+\right.\\
&&\left.-7u\f{\d^{2}s_{2}}{\d u^{2}}+15u\f{\d^{2}s_{1}}{\d u^{2}}
-15u\f{\d^{3}s_{0}}{\d
u^{3}}\right)\ux(\uxx)^{2}+\left(6s_{4}-2\f{\d s_{3}}{\d
u}-\f{\d^{2} s_{2}}{\d u^{2}}+3\f{\d^{2} s_{1}}{\d
u^{2}}-3\f{\d^{3} s_{0}}{\d u^{3}}+\right.\\
&&\left.+24u\f{\d s_{4}}{\d u}-8u\f{\d^{2} s_{3}}{\d u^{2}}
-2u\f{\d^{3} s_{2}}{\d u^{3}}+10u\f{\d^{3} s_{1}}{\d
u^{3}}-10u\f{\d^{4} s_{0}}{\d u^{4}}\right)\ux^{3}\uxxx+
\left(\f{3}{2}\f{\d s_{4}}{\d u}
-\f{1}{2}\f{\d^{2} s_{3}}{\d u^{2}}
+\f{1}{2}\f{\d^{3} s_{1}}{\d u^{3}}+\right.\\
&&\left.-\f{1}{2}\f{\d^{4} s_{0}}{\d u^{4}}
+3u\f{\d^{2} s_{4}}{\d u^{2}}
-u\f{\d^{3} s_{3}}{\d u^{3}}
+u\f{\d^{4} s_{1}}{\d u^{4}}
-u\f{\d^{5} s_{0}}{\d u^{5}}\right)\ux^{5}                             
\end{eqnarray*}
\begin{eqnarray*}
&&b_{1}=c_{0}\uxxxx+c_{1}\ux\uxxx+c_{2}(\uxx)^{2}+c_{3}\ux^{2}\uxx
+c_{4}\ux^{4}=\\
&&\left(-2s_{0}-7u\f{\d s_{0}}{\d u}+6us_{1}-6us_{2}\right)\uxxxx
+\left(-2s_{2}-s_{1}-2\f{\d s_{0}}{\d u}-2us_{3}
-12u\f{\d s_{2}}{\d u}
+16u\f{\d s_{1}}{\d u}+\right.\\
&&\left.-20u\f{\d^{2} s_{0}}{\d
u^{2}}\right)\ux\uxxx+\left(-3s_{2}-2us_{3}-8u\f{\d s_{2}}{\d 
u}+12u\f{\d
s_{1}}{\d u}-15u\f{\d^{2} s_{0}}{\d
u^{2}}\right)(\uxx)^{2}+\left(-5s_{3}+\right.\\
&&\left.-2\f{\d s_{2}}{\d u}+\f{9}{2}\f{\d s_{1}}{\d u}-6\f{\d^{2}
s_{0}}{\d u^{2}}+24us_{4}-13u\f{\d s_{3}}{\d u}-6u\f{\d^{2} s_{2}}{\d
u^{2}}+24u\f{\d^{2} s_{1}}{\d u^{2}}-30u\f{\d^{3} s_{0}}{\d
u^{3}}\right)\ux^{2}\uxx+\\
&&+\left(-3s_{4}-\f{\d s_{3}}{\d u}+\f{3}{2}\f{\d^{2} s_{1}}{\d
u^{2}}-2\f{\d^{3} s_{0}}{\d u^{3}}+6u\f{\d s_{4}}{\d u}
-3u\f{\d^{2} s_{3}}{\d u^{2}}
+4u\f{\d^{3} s_{1}}{\d u^{3}}
-5u\f{\d^{4} s_{0}}{\d u^{4}}\right)\ux^{4}
\end{eqnarray*}
\begin{eqnarray*}
&&b_{2}=d_{0}\uxxx+d_{1}\ux\uxx+d_{2}\ux^{3}=
\left(-6us_{2}+6us_{1}-10u\f{\d s_{0}}{\d u}-7s_{0}\right)\uxxx+\\
&&+\left(-6s_{2}-3s_{1}-3\f{\d s_{0}}{\d u}-6us_{3}-6u\f{\d s_{2}}{\d
u}+18u\f{\d s_{1}}{\d u}-30u\f{\d^{2} s_{0}}{\d u^{2}}\right)\ux\uxx+\\
&&\left(-3s_{3}+\f{3}{2}\f{\d s_{1}}{\d u}-3\f{\d^{2} s_{0}}{\d u^{2}}-
3u\f{\d s_{3}}{\d u}+6u\f{\d^{2} s_{1}}{\d u^{2}}-10u\f{\d^{3} s_{0}}{\d
u^{3}}\right)\ux^{3}
\end{eqnarray*}
\begin{eqnarray*}
&&b_{3}=p_{0}\uxx+p_{1}\ux^{2}=\left(-8s_{0}-4us_{2}+4us_{1}-10u\f{\d
s_{0}}{\d u}\right)\uxx+\\
&&+\left(-3s_{1}-2\f{\d s_{0}}{\d u}-2us_{3}+4u\f{\d s_{1}}{\d
u}-10u\f{\d^{2} s_{0}}{\d u^{2}}\right)\ux^{2}\\
&&b_{4}=q\ux=\left(-5s_{0}-5u\f{\d s_{0}}{\d u}\right)\ux\\
&&b_{5}=k=-2us_{0}
\end{eqnarray*}

\noindent
We are almost able to write the equation
\begin{equation}
d_{1}d_{2}X^{(4)}_{2}-\f{1}{2}\left[d_{1}X_{2}^{(2)},
d_{1}X_{2}^{(2)}\right]=0
\end{equation}

\noindent
in terms of $b_{k}$. In fact to obtain the
term $d_{1}d_{2}X_{2}^{(4)}$ it is sufficient to add the following
terms to the formula we have found before:
\begin{eqnarray*}
&&\f{1}{2}\f{\d b_{4}}{\d u}\delta^{(4)}(x-y)\delta^{(1)}(x-z)
+\f{1}{2}\f{\d b_{5}}{\d u}\delta^{(5)}(x-y)\delta^{(1)}(x-z)+\\
&&+\f{1}{2}\f{\d b_{4}}{\d \ux}\delta^{(4)}(x-y)\delta^{(2)}(x-z)
+\f{1}{2}\f{\d b_{3}}{\d \uxx}\delta^{(3)}(x-y)\delta^{(3)}(x-z)+\\
&&+\f{1}{2}\f{\d b_{2}}{\d \uxxx}\delta^{(2)}(x-y)\delta^{(4)}(x-z)
+\f{1}{2}\f{\d b_{1}}{\d \uxxxx}\delta^{(1)}(x-y)\delta^{(5)}(x-z)\\
&&+\f{1}{2}\f{\d b_{0}}{\d \uxxxxx}\delta(x-y)\delta^{(6)}(x-z)+...
\end{eqnarray*} 

\noindent
and to rewrite the terms in the usual way.                                            

\noindent
Defining the functions $f$, $g$, $h$ in the following way:
\begin{eqnarray*}
&&f:=\f{\d b_{5}}{\d u}-\f{\d b_{4}}{\d u_{x}}+\f{\d b_{2}}{\d u_{xxx}}
-\f{\d b_{1}}{\d u_{xxxx}}\\
&&g:=\f{\d b_{1}}{\d u_{xxx}}-\f{\d b_{2}}{\d u_{xxx}}+\f{\d b_{3}}{\d 
u_{x}}-\f{\d b_{4}}{\d u}-2\d_{x}\left(\f{\d b_{1}}{\d u_{xxxx}}\right)
+\d_{x}\left(\f{\d b_{2}}{\d u_{xxx}}\right)-\d_{x}\left(\f{\d b_{4}}{\d 
u_{x}}\right)+2\d_{x}\left(\f{\d b_{5}}{\d u}\right)\\
&&h:=\f{\d b_{1}}{\d u_{x}}-\d_{x}\left(\f{\d b_{1}}{\d u_{xx}}\right)
+\d_{x}^{2}\left(\f{\d b_{1}}{\d u_{xxx}}\right)-\d_{x}^{3}\left(\f{\d 
b_{1}}{\d u_{xxxx}}\right)-\f{\d b_{2}}{\d u}+\d_{x}\left(\f{\d b_{3}}{\d 
u}\right)-\d_{x}^{2}\left(\f{\d b_{4}}{\d u}\right)+\\
&&+\d_{x}^{3}\left(\f{\d b_{5}}{\d u}\right)
\end{eqnarray*}

\noindent
we have
\begin{eqnarray*}
&&d_{1}d_{2}X=2f\delta^{1}\wedge\delta^{5}+5f\delta^{2}\wedge\delta^{4}+
2g\delta\wedge\delta^{5}+(5f_{x}+5g)\delta^{1}\wedge\delta^{4}+(6f_{x}+2g)
\delta^{2}\wedge\delta^{3}+5g_{x}\delta\wedge\delta^{4}\\
&&+(4f_{xx}+8g_{x})
\delta^{1}\wedge\delta^{3}+(2h+4g_{xx})\delta\wedge\delta^{3}+(f_{xxx}
+3h+3g_{xx})\delta^{1}\wedge\delta^{2}+(3h_{x}+g_{xxx})\delta
\wedge\delta^{2}+\\
&&+h_{xx}\delta\wedge\delta^{1}
\end{eqnarray*}

\noindent
Then the equation $
d_{1}d_{2}X^{(4)}_{2}-\f{1}{2}\left[d_{1}X_{2}^{(2)},
d_{1}X_{2}^{(2)}\right]=0$ is eqivalent to the conditions 
\begin{eqnarray*}
&&f-4s\f{\d s}{\d u}=0\\
&&g=0\\
&&h=0
\end{eqnarray*}

\noindent
The first equation connects the second order deformation with the fourth 
order deformation:
\begin{eqnarray}
s_{0}=-2s\f{\d s}{\d u}
\label{equation3}
\end{eqnarray}

\noindent 
The second and the third equation give the conditions
\begin{eqnarray}
&&s_{1}-s_{2}=\f{\d s_{0}}{\d u} \label{equation4}\\
&&3s_{4}-\f{\d s_{3}}{\d u}+\f{\d^{2} s_{2}}{\d u^{2}}=0
\label{equation5}                                   
\end{eqnarray}

\noindent
\subsubsection{Trivial deformations}

\noindent
In this case trivial deformations can be written as $d_{1}A+d_{2}B$ with
\begin{eqnarray}
&&A=\int_{S^{1}}(A_{0}(u)u_{xxx}+A_{1}(u)u_{x}u_{xx}+A_{2}(u)u_{x}^{3})dx\\
&&B=\int_{S^{1}}(B_{0}(u)u_{xxx}+B_{1}(u)u_{x}u_{xx}+B_{2}(u)u_{x}^{3})dx
\end{eqnarray}

\noindent
A straightforward calculation shows that, for trivial
deformations, the coefficients $s_{i}$ are
\begin{eqnarray}
&&s_{0}=0\\
&&s_{1}=s_{2}=3\tilde{A}+3u\tilde{B}\\
&&s_{3}=6\f{\d\tilde{A}}{\d
u}+6u\f{\d\tilde{B}}{\d u}+\f{3}{2}\tilde{B}\\
&&s_{4}=\f{\d^{2}\tilde{A}}{\d u^{2}}+6u\f{\d^{2}\tilde{B}}{\d 
u^{2}}+\f{1}{2}\f{\d\tilde{B}}{\d u}
\end{eqnarray}                                                     

\noindent
with
\begin{eqnarray*}
&&\tilde{A}=2A_{2}-\f{\d A_{1}}{\d u}+\f{\d^{2} A_{0}}{\d
u^{2}}\\
&&\tilde{B}=2B_{2}-\f{\d B_{1}}{\d u}+\f{\d^{2} B_{0}}{\d u^{2}}
\end{eqnarray*}

\noindent
Now we are able to prove that $P^{(4)}_{2}$ is trivial if and only if 
$s_{0}=0$.\footnote{to avoid any confusion with the other functions 
$s_{0}$ defined in this paper, in the theorem \ref{classification} 
we 
have substituted $s_{0}$ 
for $\tilde{s}$.} 

\noindent
{\bf Proof}

\noindent
Trivial deformations satisfy equations
\begin{eqnarray*}
&&s_{0}=0\\
&&s_{1}=s_{2}\\
&&3s_{4}-\f{\d s_{3}}{\d u}+\f{\d^{2} s_{2}}{\d u^{2}}=0
\end{eqnarray*}

\noindent
In fact 
\begin{eqnarray*}
3\f{\d^{2}}{\d u^{2}}\left(\tilde{A}+\tilde{B}\right)
-\f{3}{2}\f{\d\tilde{B}}{\d u}-\f{\d}{\d u}\left(
6\f{\d}{\d
u}\left(\tilde{A}+\tilde{B}\right)-\f{9}{2}\tilde{B}\right)+\f{\d^{2}}
{\d u^{2}}\left(3\left(\tilde{A}+\tilde{B}\right)\right)=0
\end{eqnarray*}

\noindent
Conversely, we can always write a deformation $X$ (with $s=0$) as
$X=d_{1}A+d_{2}B$. It is sufficient to choose $A$ and $B$ in
such way that the following equations hold:
\begin{eqnarray*}
&&\tilde{B}=\f{4}{9}\f{\d s_{2}}{\d u}-\f{2}{9}s_{3}\\
&&\tilde{A}=\f{1}{3}s_{2}-\tilde{B}
\end{eqnarray*}
                                                        
\subsubsection{Deformations: explicit form}

\noindent
By using the formula (\ref{formula2})
\begin{eqnarray*}
&&d_{1}X=\\
&&=\sum_{s}\left(\left(\d^{s}_{y}\delta'(y-x)\right)\f{\d}{\d
u^{(s)}(y)}(s_{0}u_{yyyy}+s_{1}u_{y}u_{yyy}+s_{2}u_{yy}^{2}+s_{3}u_{y}^{4})
+\right.\\
&&\left.-\left(\d^{s}_{x}\delta'(x-y)\right)\f{\d}{\d
u^{(s)}(x)}(s_{0}u_{xxxx}+s_{1}u_{x}u_{xxx}+s_{2}u_{xx}^{2}+s_{3}u_{x}^{4})
\right)
\end{eqnarray*}

\noindent
and conditions (\ref{equation3}), (\ref{equation4}) and (\ref{equation5}) 
it is easy to get the formula

\begin{eqnarray*}
&&d_{1}X_{2}^{(4)}=-2s_{0}\delta^{(5)}(x-y)
-5\left(\d_{x}s_{0}\right)\delta^{(4)}(x-y)
-10\left(\d^{2}_{x}s_{0}\right)\delta^{(3)}(x-y)+\\
&&-10\left(\d_{x}^{3}s_{0}\right)\delta^{(2)}(x-y)
-3\left(\d_{x}^{4}s_{0}\right)\delta^{(1)}(x-y)
+2w\delta^{(3)}(x-y)+3\left(\d_{x}w\right)\delta^{(2)}(x-y)+\\
&&+\left(\d_{x}^{2}w\right)\delta^{(1)}(x-y)
\end{eqnarray*}

\noindent
whith
\begin{equation}
w=w_{0}u_{xx}+w_{1}u_{x}^{2}=2\f{\d s_{0}}{\d
u}u_{xx}+w_{1}u_{x}^{2}
\end{equation}

\noindent
where $w_{1}$ is an arbitrary function
of $u$ and $s_{0}$ is related to
the function $s$ appearing in second order deformation by the equation 
\begin{equation}
s_{0}=-2s\f{\d s}{\d u}
\end{equation}                                               

\newsection{Quasi-triviality}
\label{sec6}

\begin{de}
The group of transformations 
\begin{equation}
u\rightarrow\bar{u}=\sum_{k}\epsilon^{k}\f{F_{k}(u,u_{x},u_{xx},...)}
{G_{k}(u,u_{x},u_{xx},...)}
\end{equation}

\noindent
where $F_{k}$, $G_{k}\in A$, $deg(F_{k})-deg(G_{k})=k$ and $\f{\d 
F_{0}}{\d u}\neq 0$ is called {\bf quasi-Miura group}.
\end{de}

\noindent
A deformation is called quasi-trivial if it can be eliminated by the 
action of the quasi-Miura group.

\noindent
We have seen that, in general, the second and fourth order 
deformations are not trivial. In this section we show that they are 
quasi-trivial.

\subsection{Second order}

\begin{te}
If $X^{(2)}_{2}=d_{1}A+d_{2}B$, where $A$ and $B$ are local 
functionals whose densities are ratios of 
differential polynomials, then the secon order deformation 
$P_{\lambda}=P^{(0)}_{1}-\lambda(P^{(0)}_{2}+\epsilon^{2}d_{1}X^{(2)}_{2})$ 
is trivial.
\end{te}

\noindent
{\bf Proof}

\noindent
$X^{(2)}_{2}=d_{1}A+d_{2}B\Rightarrow P^{(2)}_{2}=-d_{2}d_{1}B$

\noindent
This implies
\begin{eqnarray*}
&&P^{(2)}_{2}=Lie_{X}P^{(0)}_{2}\\
&&Lie_{X}P^{(0)}_{1}=0
\end{eqnarray*}

\noindent
with 
\begin{equation}
\label{vector}
X=-d_{1}B
\end{equation}

\begin{te}
All second order deformations are quasi-trivial.
\end{te}

\noindent
{\bf Proof}

\noindent
If we choose
\begin{eqnarray}
&&A=\int_{S^{1}}(a(u)\f{u_{xx}}{u_{x}})dx\\
&&B=\int_{S^{1}}(b(u)\f{u_{xx}}{u_{x}})dx
\end{eqnarray}

\noindent
By using formula (\ref{formula1}) we get
\begin{eqnarray*}
&&d_{1}A+d_{2}B=\\
&&=\left(2\f{\d^{2}a}{\d u^{2}}+2u\f{\d^{2}b}{\d u^{2}}
+\f{1}{2}\f{\d b}{\d u}\right)u_{xx}-
\left(\f{\d a}{\d u}+u\f{\d b}{\d u}\right)\f{u_{xx}^{2}}{u_{x}^{2}}+\\
&&\left(\f{\d a}{\d u}+u\f{\d b}{\d u}\right)\f{u_{xxx}}{u_{x}}+
\left(\f{\d^{3}a}{\d u^{3}}+u\f{\d^{3}b}{\d u^{3}}+\f{1}{2}\f{\d^{2} b}{\d 
u^{2}}\right)u_{x}^{2}
\end{eqnarray*} 

\noindent
If we put
\begin{equation}
a(u)=-ub(u)+\int_{u_{0}}^{u}b(u)du
\end{equation}

\noindent
Then
\begin{eqnarray*}
&&\f{\d a}{\d u}=-u\f{\d b}{\d u}\\
&&\f{\d^{2} a}{\d u^{2}}=-\f{\d b}{\d u}-u\f{\d^{2} b}{\d u^{2}}\\
&&\f{\d^{3} a}{\d u^{3}}=-2\f{\d^{2} b}{\d u^{2}}-u\f{\d^{3} b}{\d u^{3}}
\end{eqnarray*}

\noindent
and these equations imply
\begin{equation}
d_{1}A+d_{2}B=-\f{3}{2}\left(\f{\d b}{\d u}u_{xx}+\f{\d^{2} b}{\d 
u^{2}}u_{x}^{2}\right)
\end{equation}

\noindent
that is equal to $X^{(2)}_{2}=s(u)u_{xx}+\f{\d s}{\d u}u_{x}^{2}$ 
if we choose 
\begin{equation}
\label{choice}
b(u)=-\f{2}{3}\int_{u_{0}}^{u}s(u)du
\end{equation}

\subsection{Fourth order}
\newtheorem{lem}{Lemma}
\begin{lem}
A fourth order deformation is quasi-trivial if and only if there exists a 
vector field $Y$ such that

\begin{eqnarray}
&&Lie_{Y}P^{(0)}_{1}=0\\
&&P^{(4)}_{2}-\f{1}{2}Lie_{X}^{2} P^{(0)}_{2}=Lie_{Y}P^{(0)}_{2}
\end{eqnarray}

\noindent
where $X$ is the vector field (\ref{vector}).
\end{lem} 

\noindent
{\bf Proof}

\noindent
The reduction of the fourth order deformation to the form 
$P^{(0)}_{1}-\lambda P^{(0)}_{2}$ can be achieved in two steps.

\noindent
In the first step one kills the second order part of the deformation 
 $P^{(0)}_{2}$ with the quasi-Miura transformation generated by the vector 
field  (\ref{vector}):
\begin{equation*}
P^{(0)}_{1}-\lambda\left(P^{(0)}_{2}+\epsilon^{2}P^{(2)}_{2}
+\epsilon^{4}P^{(4)}_{2}+O(\epsilon^{5})\right)
\rightarrow 
P^{(0)}_{1}-\lambda\left(P^{(0)}_{2}+\epsilon^{4}
\left(P^{(4)}_{2}-\f{1}{2}Lie_{X}^{2}P^{(0)}_{2}
+O(\epsilon^{5})\right)\right). 
\end{equation*}

\noindent
In the second step one kills the fourth order part of the deformation
$P^{(4)}_{2}-\f{1}{2}Lie_{X}^{2}P^{(0)}_{2}$ with the quasi-Miura 
transformation generated by the vector field $Y$.

\begin{te}
All fourth order deformations are quasi-trivial.
\end{te}

\noindent
To prove the theorem we need the following
\begin{lem}
If $X^{(4)}_{2}=d_{1}\tilde{A}+d_{2}\tilde{B}-\f{1}{2}[d_{1}B,d_{2}B]$
where 
\begin{equation*}
B=\int_{S^{1}}b(u)\f{u_{xx}}{u_{x}}dx
\end{equation*}

\noindent
and $b(u)$ is given by the expression (\ref{choice}), then the fourth 
order deformation 
\begin{equation*}
P^{(0)}_{1}-\lambda\left(P^{(0)}_{2}+\epsilon^{4}
\left(P^{(4)}_{2}-\f{1}{2}Lie_{X}^{2}P^{(0)}_{2}\right)\right)
\end{equation*}

\noindent
is trivial.
\end{lem}     

\noindent
{\bf Proof}

\noindent
$P^{(4)}_{2}=d_{1}X^{(4)}_{2}
=d_{1}d_{2}\tilde{B}-d_{1}\f{1}{2}[d_{1}B,d_{2}B]$

\noindent
By using the graded Jacobi identity we get 
\begin{equation}
P^{(4)}_{2}-\f{1}{2}Lie_{X}^{2} P^{(0)}_{2}=Lie_{Y}P^{(0)}_{2} 
\end{equation}

\noindent
where $X=-d_{1}B$ and $Y=-d_{1}\tilde{B}$. Moreover
\begin{equation*}
Lie_{Y}P^{(0)}_{1}=0
\end{equation*} 

\noindent
Then to prove the theorem it is sufficient to show that the vector field 
$X^{(4)}_{2}=s_{0}u_{xxxx}+s_{1}u_{x}u_{xxx}+s_{2}u_{xx}^{2}
+s_{3}u_{x}^{2}u_{xx}+s_{4}
u_{x}^{4}$ \footnote{We recall that the coefficients of the 
vector field $X^{(4)}_{2}$ satisfy the condition (\ref{equation3}), 
(\ref{equation4}) and (\ref{equation5})} can be written  in the 
form
\begin{equation*}
X^{(4)}_{2}=d_{1}\tilde{A}+d_{2}\tilde{B}-\f{1}{2}[d_{1}B,d_{2}B] 
\end{equation*}

\noindent
We start calculating the term $-\f{1}{2}[d_{1}B,d_{2}B]$. In this case the 
Schouten bracket coincides with the commutator of the vector fields 
\begin{eqnarray*}
d_{1}B=2\f{\d^{2} b}{\d u^{2}}u_{xx}-\f{\d b}{\d 
u}\f{u_{xx}^{2}}{u_{x}^{2}}+\f{\d b}{\d u}\f{u_{xxx}}{u_{x}}+\f{\d^{3} 
b}{\d u^{3}}u_{x}^{2}
\end{eqnarray*}

\noindent
and
\begin{eqnarray*}
d_{2}B=ud_{1}B+\f{1}{2}\left(\f{\d b}{\d u}u_{xx}+\f{\d^{2} b}{\d 
u^{2}}u_{x}^{2}\right)
\end{eqnarray*}

\noindent
The result is

\begin{eqnarray*}
&&\nonumber -\f{1}{2}[d_{1}B,d_{2}B]^{x}=-\f{1}{2}\sum_{s}
\left(\d_{x}^{s}(d_{1}B)\f{\d d_{2}B}{\d u^{(s)}(x)}
-\d_{x}^{s}(d_{2}B)\f{\d d_{1}B}{\d u^{(s)}(x)}\right)=\\ 
&&\nonumber u_{xxxx}\left(\f{33}{4}
\f{\d b}{\d u}\f{\d^{2} b}{\d u^{2}}\right)+
u_{x}u_{xxx}
\left(\f{27}{4}\left(\f{\d^{2} b}{\d u^{2}}\right)^{2}+
\f{45}{4}\f{\d b}{\d u}\f{\d^{3} b}{\d u^{3}}\right)+\\
&&\nonumber +u_{xx}^{2}
\left(-\f{3}{4}\left(\f{\d^{2} b}{\d u^{2}}\right)^{2}+
\f{15}{4}\f{\d b}{\d u}\f{\d^{3} b}{\d u^{3}}\right)      
+u_{x}^{2}u_{xx}
\left(\f{45}{4}\f{\d^{2} b}{\d u^{2}}\f{\d^{3} b}{\d u^{3}}+
\f{21}{2}\f{\d b}{\d u}\f{\d^{4} b}{\d u^{4}}\right)+\\
&&\nonumber +u_{x}^{4}
\left(\f{9}{4}\f{\d^{2} b}{\d u^{2}}\f{\d^{4} b}{\d u^{3}}+  
\f{3}{2}\f{\d^{5} b}{\d u^{5}}\f{\d b}{\d u}+
\f{3}{4}\left(\f{\d^{3} b}{\d u^{3}}\right)^{2}\right)+
\f{9}{2}\f{u_{xxx}^{2}}{u_{x}^{2}} 
\left(\f{\d b}{\d u}\right)^{2}+\\
&&\nonumber -9\f{u_{xx}^{4}}{u_{x}^{4}}
\left(\f{\d b}{\d u}\right)^{2} 
+2\f{u_{xxxxx}}{u_{x}}
\left(\f{\d b}{\d u}\right)^{2}
-\f{u_{xx}u_{xxx}}{u_{x}}
\left(\f{27}{2}\f{\d b}{\d u}\f{\d^{2} b}{\d u^{2}}\right)+\\
&&+\f{u_{xx}^{3}}{u_{x}^{2}}
\left(\f{33}{4}\f{\d b}{\d u}\f{\d^{2} b}{\d u^{2}}\right)+ 
18\f{u_{xx}^{2}u_{xxx}}{u_{x}^{3}}
\left(\f{\d b}{\d u}\right)^{2}
-6\f{u_{xx}u_{xxxx}}{u_{x}^{2}}
\left(\f{\d b}{\d u}\right)^{2} 
\label{term}
\end{eqnarray*} 

\noindent
The problem now is to guess the form of the local functionals $\tilde{A}$ 
and $\tilde{B}$. 
We know that the degree of  $\tilde{A}$  and $\tilde{B}$ is equal to 3 but 
all ratios of differential polynomials of the form 
\begin{equation*}
\f{P(u,u_{x},u_{xx},...)}{Q(u,u_{x},u_{xx},...)}
\end{equation*}

\noindent
where $deg(P)-deg(Q)=3$, are allowed. 

\noindent
The form of the coefficients in (\ref{term}) suggests that $\tilde{A}$  
and $\tilde{B}$ contain only terms of the form
\begin{equation*} 
\f{P(u,u_{x},u_{xx}...)}{u_{x}^{i}}
\end{equation*}

\noindent
for some integer $i$.

\noindent
Let us try with the local functionals
\begin{eqnarray*}
&&\tilde{a}=\int_{s^{1}}\left(h1(u)\f{u_{xx}^{2}}{u_{x}}+
+h2(u)\f{u_{xx}^{3}}{u_{x}^{3}}+h3(u)\f{u_{xx}u_{xxx}}{u_{x}^{2}}+
h4(u)\f{u_{xxxx}}{u_{x}}+
h5(u)\f{u_{xx}^{2}u_{xxx}}{u_{x}^{4}}+h6(u)\f{u_{xx}^{4}}{u_{x}^{5}}+\right.\\
&&\left.+h7(u)\f{u_{xx}u_{xxxx}}{u_{x}^{3}}+
h8(u)\f{u_{xxx}^{2}}{u_{x}^{3}}+h9(u)\f{u_{xxxxx}}{u_{x}^{2}}+
h10(u)\f{u_{xxx}u_{xxxx}}{u_{x}^{4}}+
h11(u)\f{u_{xxxx}^{2}}{u_{x}^{5}}\right)dx\\                                
&&\tilde{b}=\int_{s^{1}}\left(k1(u)\f{u_{xx}^{2}}{u_{x}}+
+k2(u)\f{u_{xx}^{3}}{u_{x}^{3}}+k3(u)\f{u_{xx}u_{xxx}}{u_{x}^{2}}+
k4(u)\f{u_{xxxx}}{u_{x}}+
k5(u)\f{u_{xx}^{2}u_{xxx}}{u_{x}^{4}}+k6(u)\f{u_{xx}^{4}}{u_{x}^{5}}+\right.\\
&&\left.+k7(u)\f{u_{xx}u_{xxxx}}{u_{x}^{3}}+
k8(u)\f{u_{xxx}^{2}}{u_{x}^{3}}+k9(u)\f{u_{xxxxx}}{u_{x}^{2}}+
k10(u)\f{u_{xxx}u_{xxxx}}{u_{x}^{4}}+
k11(u)\f{u_{xxxx}^{2}}{u_{x}^{5}}\right)dx
\end{eqnarray*}  

\noindent
In order to determine the exact form of the coefficients $h_{i}$ 
and $k_{i}$ one 
has just to compare $d_{1}\tilde{a}+d_{2}\tilde{b}$ with 
$\f{1}{2}[d_{1}B,d_{2}B]$.

\noindent
The first step is to kill the coefficients of rational terms in 
$d_{1}\tilde{a}+d_{2}\tilde{b}$ that don't appear in 
$-\f{1}{2}[d_{1}B,d_{2}B]$ and the second step is to eliminate the 
remaining 
rational terms. 

\noindent
The calculations are very long, we give here only the final 
result:
\footnote{In fact there is a certain freedom in
the choice of coefficients $h_{i}$ and $k_{i}$; for example it is 
not
necessary to put $k4(u)$ equal to 0.}                               
\begin{eqnarray*}
&&h_{i}(u)=-uk_{i}(u),\ i=1,...,11\\
&&k1(u)=-\f{1}{28}k8(u)-\f{15}{56}k2(u)\\
&&k3(u)=4k1(u)\\
&&k4(u)=0\\
&&k5(u)=-6k8(u)\\
&&k6(u)=4k8(u)\\
&&k7(u)=k8(u)\\
&&k9(u)=k10(u)=k11(u)=0\\
&&k2(u)+\f{\d k8}{\d u}=\f{7}{5}\left(\f{\d b}{\d u}\right)^{2}
\end{eqnarray*}

\noindent
Whith this choice and taking into account (\ref{choice}), by using 
formula 
(\ref{formula1}) we get

\begin{eqnarray*}
&&-\f{1}{2}[d_{1}B,d_{2}B]+d_{1}\tilde{a}+d_{2}\tilde{b}=\\
&&-2s\f{\d s}{\d u}u_{xxxx}
+\left(-\f{7}{3}s\f{\d^{2} s}{\d 
u^{2}}-\f{13}{3}\left(\f{\d s}{\d u}\right)^{2}\right)u_{x}u_{xxx}  
+\left(-\f{1}{3}s\f{\d^{2} s}{\d
u^{2}}-\f{7}{3}\left(\f{\d s}{\d u}\right)^{2}\right)u_{xx}^{2}+\\
&&+\left(\f{5}{3}s\f{\d^{3} s}{\d
u^{3}}-4\f{\d s}{\d u}\f{\d^{2} s}{\d u^{2}}\right)u_{x}^{2}u_{xx}
+\left(\f{1}{3}\left(\f{\d^{2} s}{\d
u^{2}}\right)^{2}+\f{\d s}{\d u}\f{\d^{3} s}{\d 
u^{3}}+\f{2}{3}s\f{\d^{4} s}{\d u^{4}}\right)u_{x}^{4}=\\
&&s_{0}'u_{xxxx}+s_{1}'u_{x}u_{xxx}+s_{2}'u_{xx}^{2}
+s_{3}'u_{x}^{2}u_{xx}+s_{4}'u_{x}^{4}
\end{eqnarray*}                                                           
           
\noindent
First of all we observe that  $s_{0}'$ is equal to 
$s_{0}$. Moreover the coefficients $s'_{i}$ satisfy equations 
(\ref{equation4}) and (\ref{equation5}):
\begin{eqnarray*}
&&s_{1}'-s_{2}'=\f{\d}{\d u}\left(-2s\f{\d s}{\d u}\right)\\
&&3s_{4}'-\f{\d s_{3}'}{\d u}+\f{\d^{2} s_{2}'}{\d u^{2}}=
\left(\f{\d^{2} s}{\d u^{2}}\right)^{2}+3\f{\d s}{\d u}\f{\d^{2} 
s}{\d u^{2}}+2s\f{\d^{4} s}{\d u^{4}}-\f{5}{3}s\f{\d^{4} s}{\d 
u^{4}}+\\ 
&&+\f{7}{3}\f{\d s}{\d u}\f{\d^{3} s}{\d u^{3}}
+4\left(\f{\d^{2} s}{\d u^{2}}\right)^{2}-\f{16}{3}\f{\d s}{\d 
u}\f{\d^{3} s}{\d u^{3}}-5\left(\f{\d^{2} s}{\d 
u^{2}}\right)^{2}-\f{1}{3}s\f{\d^{4} s}{\d u^{4}}=0
\end{eqnarray*}

\noindent
Then the vector field
\begin{eqnarray*}
&&X-(-\f{1}{2}[d_{1}B,d_{2}B]+d_{1}\tilde{a}+d_{2}\tilde{b})=\\
&&=(s_{0}-s_{0}')u_{xxxx}+(s_{1}-s_{1}')u_{x}u_{xxx}+(s_{2}-s_{2}')u_{xx}^{2}
+(s_{3}-s_{3}')u_{x}^{2}u_{xx}+(s_{4}-s_{4}')u_{x}^{4}
\end{eqnarray*}

\noindent
is trivial. Indeed:
\begin{eqnarray*}
s_{0}-s_{0}'=0
\end{eqnarray*}

\noindent
Moreover the equations
\begin{eqnarray*}
&&s_{1}-s_{2}=\f{\d s_{0}}{\d u}\\ 
&&s_{1}-s_{2}=\f{\d s_{0}}{\d u} 
\end{eqnarray*} 

\noindent
imply
\begin{eqnarray*} 
s_{1}-s_{1}'=s_{2}-s_{2}'
\end{eqnarray*} 

\noindent
Finally
\begin{eqnarray*}
&&3(s_{4}-s_{4}')-\f{\d}{\d u}(s_{3}-s_{3}')+\f{\d^{2}}{\d 
u^{2}}(s_{2}-s_{2}')=\\
&&3s_{4}-\f{\d}{\d u}s_{3}+\f{\d^{2}}{\d u^{2}}s_{2}
-\left(3s_{4}'-\f{\d}{\d u}s_{3}'+\f{\d^{2}}{\d u^{2}}s'_{2}\right)=0                                   
\end{eqnarray*}

\noindent
This means that
\begin{equation*}
X-(-\f{1}{2}[d_{1}B,d_{2}B]+d_{1}\tilde{a}+d_{2}\tilde{b})= 
d_{1}a+d_{2}b
\end{equation*}

\noindent
that is

\begin{equation*}
X=-\f{1}{2}[d_{1}B,d_{2}B]+d_{1}(\tilde{a}+a)+d_{2}(\tilde{b}+b)
\end{equation*}

\newsection{Conclusions}

\noindent
In this paper we have studied the problem of classification of  
deformations of bihamiltonian structure of hydrodynamic type.

\noindent
The main result of the paper is that, up to the fourth order, any 
deformation can be reduced to the form (\ref{class}) depending only on a 
functional 
parameter $s(u)$.

\noindent
These deformations give rise to an infinite hierarchy of 
``almost-commuting'' hamiltonian equations.

\noindent
In this paper we started studying numerically one of the equations of the 
deformed hierarchy corresponding to second order deformations.

\noindent
The results obtained indicate the existence of the analogue 
of 2-soliton solutions 
at least for small times and for small amplitudes, but a deeper analysis 
is still necessary.\\ 

\noindent
We have seen that one of the consequence of classification theorem
is that deformations of the Magri bihamiltonian structure are trivial up 
to fourth order. This fact suggests that in our classification problem the
Magri bihamiltonian structure is a stable object  according to the 
definition of Arnold.
\footnote{A classification problem can be thought as a 
decomposition of a  space of objects in equivalence classes (see 
\cite{Arnold}); 
in our 
case the objects are deformations; two deformations are equivalent if 
and only if they can be reduced to the same form by the action of 
Miura group. An object is called stable if a sufficiently 
small ``neighbourhood'' 
of this object contains only objects of the same 
class.}
It would be interesting to investigate if 
this situation is more general, that is, if completely integrable systems 
of bihamiltonian type are the stable objects of a corresponding 
classification problem. For example it is possible to apply the same 
techniques 
used in this paper to the ``symplectic'' case where the leading term in 
the parameter $\epsilon$ has the form

\begin{eqnarray}
P_{1,2}=h_{1,2}^{ij}\delta(x-y) 
\end{eqnarray}

\noindent
with 
\begin{eqnarray*}
&&
h_{1}^{ij}=\begin{pmatrix}
0&1&0&0&...&0&0\cr
-1&0&0&0&...&0&0\cr
0&0&0&1&...&0&0\cr
0&0&-1&0&...&0&0\cr
.&.&.&.&.&.&.\cr
.&.&.&.&.&.&.\cr 
.&.&.&.&.&.&.\cr 
0&0&0&0&...&0&1\cr
0&0&0&0&...&-1&0
\end{pmatrix}
\end{eqnarray*}

\noindent
and

\begin{eqnarray*}
&&
h_{2}^{ij}=\begin{pmatrix}
0&u_{1}&0&0&...&0&0\cr
-u_{1}&0&0&0&...&0&0\cr
0&0&0&u_{2}&...&0&0\cr
0&0&-u_{2}&0&...&0&0\cr                                                         
.&.&.&.&.&.&.\cr
.&.&.&.&.&.&.\cr                                                            
.&.&.&.&.&.&.\cr
0&0&0&0&...&0&u_{n}\cr
0&0&0&0&...&-u_{n}&0
\end{pmatrix}
\end{eqnarray*}

\noindent
Also in this case the the cohomology groups $H^{1}$ and $H^{2}$ 
associated to the differentials $d_{P_{1}}$ and $d_{P_{2}}$ are 
trivial (see \cite{Dubrovin}) and then the non trivial deformations are 
classified 
by the group 
\begin{equation*}
\f{ker\left(d_{1}d_{2}\right)}{\left(Im(d_{1})+Im(d_{2})\right)}
\end{equation*}\\

\noindent
Finally we observe that the result about quasi-triviality suggests that 
this property is a consequence of the definition of deformation and it is 
not an additional constraint.\\

\noindent
{\bf Aknowledgements}  

\noindent
I am indebted to Boris Dubrovin who proposed to me the 
classification problem and constantly helped  me with useful suggestions.
I also indebted to Arkadi Shagalov who suggested to me to use a 
two-step Lax-Landroff scheme to make numerical experiments and 
to Gregorio Falqui for useful discussions.

\end{document}